\documentclass[%
reprint,
superscriptaddress,
amsmath,amssymb,
aps,
pra,
]{revtex4-2}
\usepackage{amssymb}
\makeatletter

\newcommand{\Rmnum}[1]{\expandafter\@slowromancap\romannumeral #1@}
\makeatother
\usepackage{multirow}
\usepackage{booktabs}
\usepackage{floatflt}
\usepackage{ragged2e}
\renewcommand{\justify}{\leftskip=0pt \rightskip=0pt plus 0cm}
\usepackage{makecell}
\usepackage{floatrow}
\usepackage{graphicx}
\usepackage[label font=bf,labelformat=simple]{subfig}
\usepackage{pgfplots}
\pgfplotsset{compat=1.7}
\usepackage{amsmath}
\newtheorem{theorem}{Theorem}
\usepackage{amssymb}
\usepackage{tikz}
\usepackage{bbm}
\usepackage{dcolumn}
\usepackage{bm}
\usepackage{blindtext}
\usepackage{listings}
\usepackage{physics}
\usepackage{stmaryrd}
\usepackage[colorlinks,linkcolor=blue,anchorcolor=blue,citecolor=blue]{hyperref}
\makeatletter
\newif\if@restonecol
\makeatother

\usepackage[ruled,linesnumbered,vlined]{algorithm2e}
\usepackage{algpseudocode}

\begin{document}
\title{Facilitating Practical Fault-tolerant Quantum Computing Based on Color Codes}

\author{Jiaxuan Zhang}
\affiliation{Key Laboratory of Quantum Information, Chinese Academy of Sciences, School of Physics, University of Science and Technology of China, Hefei, Anhui, 230026, P. R. China}
\affiliation{CAS Center For Excellence in Quantum Information and Quantum Physics, University of Science and Technology of China, Hefei, Anhui, 230026, P. R. China}
\affiliation{Hefei National Laboratory, University of Science and Technology of China, Hefei 230088, P. R. China}

\author{Yu-Chun Wu}
\email{wuyuchun@ustc.edu.cn}
\affiliation{Key Laboratory of Quantum Information, Chinese Academy of Sciences, School of Physics, University of Science and Technology of China, Hefei, Anhui, 230026, P. R. China}
\affiliation{CAS Center For Excellence in Quantum Information and Quantum Physics, University of Science and Technology of China, Hefei, Anhui, 230026, P. R. China}
\affiliation{Hefei National Laboratory, University of Science and Technology of China, Hefei 230088, P. R. China}
\affiliation{Institute of Artificial Intelligence, Hefei Comprehensive National Science Center, Hefei, Anhui, 230088, P. R. China}

\author{Guo-Ping Guo}
\affiliation{Key Laboratory of Quantum Information, Chinese Academy of Sciences, School of Physics, University of Science and Technology of China, Hefei, Anhui, 230026, P. R. China}
\affiliation{CAS Center For Excellence in Quantum Information and Quantum Physics, University of Science and Technology of China, Hefei, Anhui, 230026, P. R. China}
\affiliation{Hefei National Laboratory, University of Science and Technology of China, Hefei 230088, P. R. China}
\affiliation{Institute of Artificial Intelligence, Hefei Comprehensive National Science Center, Hefei, Anhui, 230088, P. R. China}
\affiliation{Origin Quantum Computing Hefei, Anhui 230026, P. R. China}
\date{\today}
\begin{abstract}
Color code is a promising topological code for fault-tolerant quantum computing. Insufficient research on the color code has delayed its practical application. In this work, we address several key issues to facilitate practical fault-tolerant quantum computing based on color codes. First, by introducing decoding graphs with  error-rate-related weights, we obtained the threshold of $0.47\%$ of the 6,6,6 triangular color code under the standard circuit-level noise model, narrowing the gap to that of the surface code. Second, our work firstly investigates the circuit-level decoding of color code lattice surgery, and gives an efficient decoding algorithm, which is crucial for performing logical operations in a quantum computer with two-dimensional architectures. Lastly, a new state injection protocol of the triangular color code is proposed, reducing the output magic state error rate in one round of 15 to 1 distillation by two orders of magnitude compared to a previous rough protocol. We have also proven that our protocol offers the lowest logical error rates for state injection among all possible CSS codes.
\end{abstract}
\maketitle
\section{Introduction}\label{Introduction}
Quantum computation provides a potential way to solve classically intractable problems, such as integer factorization \cite{shor1999polynomial} and simulation of large quantum systems \cite{feynman1985quantum,freedman2002simulation}. To enable the practical large-scale quantum computation, quantum error correction (QEC) is the crucial technique that protects
information against the noise by encoding physical qubits into logical qubits of a certain
type of QEC code \cite{preskill1998reliable,nielsen2010quantum,terhal2015quantum}.

Much of the current work focuses on a subset of QEC codes, the stabilizer code \cite{gottesman1997stabilizer}. As a type of the stabilizer code, topological codes \cite{dennis2002topological,pachos2012introduction} including color codes \cite{bombin2006topological,kubica2018abcs} and surface codes \cite{fowler2012surface} draw extra attention since they are compatible with the real hardware limited by the local constraints in the two-dimensional (2-D) architecture. Compared to surface code, the advantage of color code is that it encodes a logical qubit with fewer physical qubits and can implement logical Clifford gates transversally.

An important indicator of the performance of a QEC code  is the threshold. However, it had long been believed that the circuit-level threshold of color codes is about an order of magnitude lower than that of surface codes, which is a significant reason why color codes fall behind in the competition with surface codes \cite{landahl2014quantum,chamberland2020triangular,fowler2012towards,stephens2014fault}. In the first part of our work, by improving the decoding algorithm, we obtain the  threshold of $0.47\%$ for the 6,6,6 triangular color code \cite{landahl2011fault}, basically the same order of magnitude as the threshold of surface codes of $0.7\%\sim0.9\%$ using an efficient decoder \cite{fowler2012towards,stephens2014fault}. 

When performing logical operations on 2-D hardware, lattice surgery \cite{horsman2012surface,landahl2014quantum,thomsen2022low,litinski2018lattice,litinski2019game} is currently the mainstream scheme since it retains locality constraints and offers lower overhead compared to braiding schemes \cite{fowler2018low}. Lattice surgery implements logical gates, including Clifford gates and non-Clifford gates, through fault-tolerant measurements of multi-body logical Pauli operators. Recent work shows that lattice surgery of color codes can further reduce the time cost by measuring an arbitrary pair of commuting logical Pauli operators in parallel \cite{thomsen2022low}, which offers additional evidence of the advantage of color codes. Moreover, in the lattice surgery of surface codes, measurements involving Pauli $Y$ operators typically needs to introduce twist defects, which increase the requirement of
device connectivity and resource costs \cite{chamberland2022circuit,gidney2023inplace}. In spite of Ref.~\cite{chamberland2022universal} that proposes twist-free lattice surgery scheme, it still requires additional space and time overhead. While in color code lattice surgery, the difficulties of measuring $Y$-type logical Pauli operators are almost non-existent, since they can transform into $X$ or $Z$-type Pauli operators through transversal single-qubit Clifford gates. Although color code lattice surgery is theoretically feasible and may have advantages, decoding strategy at the circuit level are still lacking. Our work investigates this process and proposes an efficient circuit-level decoding algorithm for a class of color code lattice surgery.

Magic states, as the ancilla states for performing non-Clifford gates \cite{bravyi2005universal}, represent another widely studied research focus in fault-tolerant quantum computing.  Obtaining high-fidelity magic states requires an expensive process called magic state distillation \cite{bravyi2012magic,campbell2017unified,litinski2019magic}. The first step of magic state distillation is injecting a physical magic state into a logical state. The quality of the initial logical magic states from state injection will remarkably affect the error rate of the output state after distillation. Several previous literatures have demonstrated that magic states from a surface code state injection protocol have a better fidelity than the operations used to construct them \cite{li2015magic,lao2022magic,singh2022high}. In contrast, well-designed state injection protocols for color codes and error rate analyses of them are still quite limited. This work proposes a new state injection protocol of color code. Compared to a rough protocol used in Ref.~\cite{beverland2021cost}, our protocol reduces the output magic state error rate in one round of 15 to 1 distillation by two orders of magnitude.

Overall, there are three main results in this paper. First, a decoding algorithm has been proposed, achieving a threshold of 0.47\% for 6,6,6 triangular color codes. Based on the projection decoder \cite{delfosse2014decoding,beverland2021cost}, we construct new decoding graphs by an automated procedure and introduce error-rate-related weights to implement more accurate matching by minimum weight perfect matching (MWPM) algorithm \cite{edmonds1965paths,higgott2022pymatching}. Our algorithm significantly improves the decoding accuracy of the previous 6,6,6 triangular color code decoder \cite{beverland2021cost} under circuit-level noise. The numerical results show that the threshold under circuit-level noise model is around $0.47\%$, which, to the best of our knowledge, is the highest threshold among all types of 2D color codes without circuit optimization. We also note that in Ref.~\cite{gidney2023new}, a higher threshold for color codes is demonstrated. However, this is primarily due to their optimization of the stabilizer measurement circuits rather than the decoder itself, and thus cannot be directly compared to our results. Second, we demonstrated the decoding process of lattice surgery between two logical qubits under circuit-level noise model. This algorithm is applicable to lattice surgery schemes on color code with colored boundaries \cite{kesselring2024anyon}. We simulate an example in which logical operators $X_{L}\otimes X_{L}$ and $Z_{L}\otimes Z_{L}$ are measured in parallel and find that the space-like effective code distance is slightly less than that the time-like one and the space-like error is the dominated logical error in color code lattice surgery. Lastly, we investigate state injection of color code and give a protocol based on post-selection. The performance of our protocol is superior to the existing state injection protocols of surface codes in terms of logical error rates, post-selection success rates, and process complexity. Furthermore, it has been proven that the logical error rate of this protocol is lowest compared to any state injection protocol among all possible CSS codes \cite{calderbank1996good,steane1996multiple}. We also  discuss how to design a proper post-selection scheme by the correlation coefficients of syndrome changes and the occurrence of logical errors.

The remaining paper is organized as follows. Sec~\ref{Preliminaries} reviews some basic but important concepts in fault-tolerant quantum computation and specifies notations used in this paper. Sec~\ref{Color code decoding} introduces the improved color code decoding strategy and presents the numerical results of the threshold. In Sec~\ref{Lattice surgery}, we describe the circuit-level decoding of color code lattice surgery and discuss its applicability. Then in Sec~\ref{State injection}, the state injection protocol of color codes is proposed, where we use a theorem to summarize its optimality. Finally, the conclusion and outlook are presented in Sec~\ref{Conclusion and outlook}.

\section{Preliminaries}\label{Preliminaries}
In this section we briefly review some important background materials for 2-D color codes, which are referred throughout the paper. Sec~\ref{basics of triangular color code} introduces the basic definitions and notations about 6,6,6 triangular color codes. In Sec~\ref{color code lattice surgery and Pauli-based computation}, we discuss the Pauli-based computation and how the lattice surgery performs in the 2-D color codes. In Sec~\ref{magic state distillation}, we review another key problem of the fault-tolerant quantum computation – magic state distillation. Lastly, the details of the circuit-level noise model are presented in Sec~\ref{circuit-level noise model}, which is the premise of the following discussions and numerical results.

\subsection{Triangular color codes and notations}\label{basics of triangular color code}
\begin{figure*}[htb]
\centering
\includegraphics[width=17.5cm]{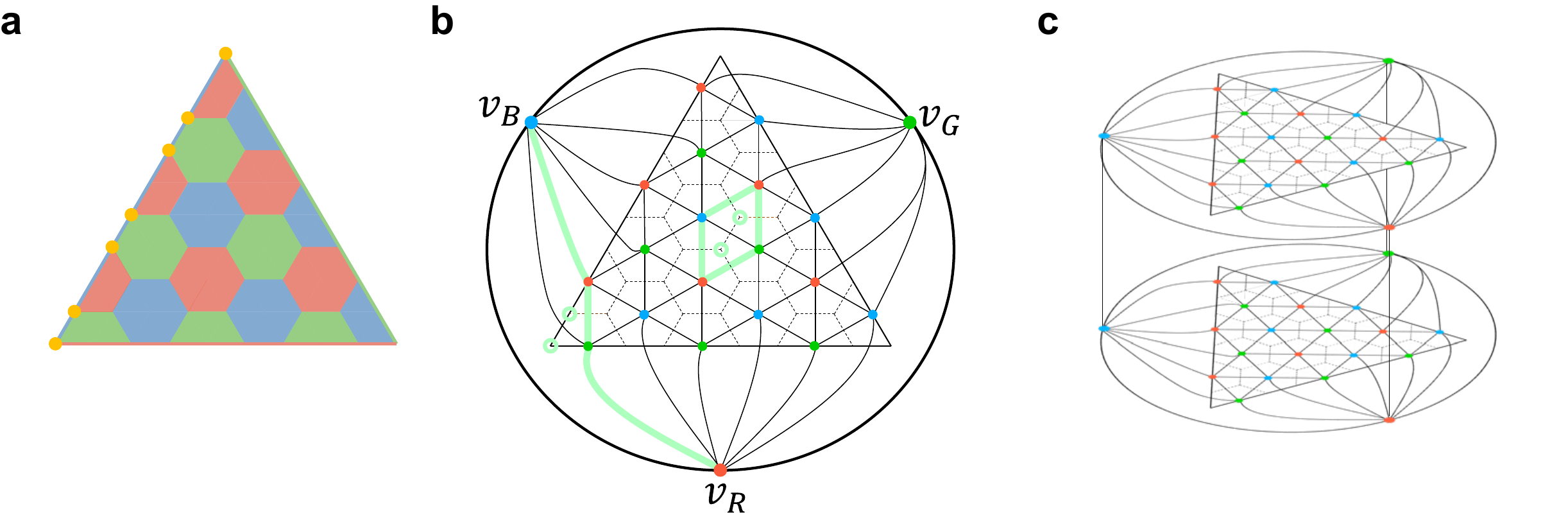}
\caption{\justify (a) Primal lattice $\mathcal{L}$ of the triangular color code. Each hexagonal or half-hexagonal face corresponds to two stabilizers which are the tensor products of $X$ and $Z$ supported by the qubits located in the vertices of the face, respectively. A representative of the $X$ (or $Z$) logical operator is defined as tensor products of $X$ (or $Z$) supported by the qubits located in a string (colored by yellow). (b) Dual lattice $\mathcal{L}^*$ of the triangular color code with $d=7$. The green lines and circles are two examples of paths and corresponding corrections in the decoding. (c) 3-D lattice $\mathcal{L}^*\times t$ (here $t=2$). Each layer of $\mathcal{L}^*\times t$ is the 2-D dual lattice $\mathcal{L}^*$. The corresponding vertices of adjacent layers are connected vertically. To avoid clutter, we only show the vertical connections of three boundary vertices.}\label{fig1}
\end{figure*}

2-D color codes are topological QEC codes constructed in a three-colorable and trivalent lattice. In this paper, we focus on the hexagonal color code with triangular boundaries (referred to as the triangular color code below). The triangular color code is defined in the hexagonal lattice $\mathcal{L}$ with three boundaries. Let us denote the sets of vertices, edges and faces of a lattice (or a graph) $\mathcal{L}$ by $V(\mathcal{L})$, $E(\mathcal{L})$ and $F(\mathcal{L})$, respectively. As inllustrated in Fig.~\ref{fig1}a, each data qubit of the triangular color code is placed in a vertex $v \in V(\mathcal{L})$, and each face $f \in F(\mathcal{L})$ corresponds to two stabilizer generators $S^X_f=\otimes_{v {\in} f} X_v$ and $S^Z_f=\otimes_{v\in f} Z_v$, respectively. The logical code space is the simultaneous eigenspace of all the stabilizers and the measurement results of all stabilizer generators form a binary string called syndrome. The triangular color code encodes one logical qubit, a representative logical Pauli operators of which can be defined as the tensor product of Pauli operators supported by a string in the boundary. The minimum weight of all logical $X$ or $Z$ operators is defined as the code distance, denoted by $d$. Note that color codes are Calderbank-Shor-Steane stabilizer codes and self-dual, i.e., the supports of $S^X_f$ and $S^Z_f$ are the same. Hence, the $X$ or $Z$ errors can be corrected separately in an analogous manner. 

Given the primal lattice $\mathcal{L}$ of the color code, it is convenient to discuss the decoding problem on its dual lattice $\mathcal{L^*}$. In the dual lattice $\mathcal{L^*}$, each face corresponds to a data qubit and each vertex corresponds two stabilizer generators $S^X_f$ and $S^Z_f$ except for three boundary vertices $v_{R}$, $v_{G}$ and $v_{B}$ (see Fig.~\ref{fig1}b). We use $F_1\subseteq F(\mathcal{L^*})$ to denote the set of faces identified with qubits affected by $X$ (or $Z$) errors and $\sigma\subseteq V(\mathcal{L^*})$ to denote the vertex set in which the identified stabilizer generators anticommutes with the Pauli error supported by ${F_1}$. 

We also need to define three restricted lattices $\mathcal{L}^*_{C}$, where $C\in \{RG,RB,GB\}$. The restricted lattice, say $\mathcal{L}^*_{RG}$, is obtained from $\mathcal{L^*}$ by removing all blue vertices of $\mathcal{L^*}$ as well as all the edge and faces incident to the removed vertices. 

Moreover, for color code decoding under the circuit-level noise model, 3-D lattices or graphs are usually employed, where the vertical direction represents the time layers, and each time layer corresponds to a 2-D lattice. We use $v_i^{t}$ to denote the vertex in a 3-D lattice or graph, where $t\in \mathbb{Z}^+$ is the number of layers in which $v_i^{t}$ is located, and $v_i$ is the projection of $v_i^{t}$ on the 2-D lattice. One type of 3-D lattice that we frequently use is $\mathcal{L^*}\times n$, which is defined by $V(\mathcal{L^*}\times n)=\{v_i^{t}|v_i\in V(\mathcal{L^*}),t=1,2,...,n\}$ and $E(\mathcal{L^*}\times n)=\{(v_i^{t},v_j^{t})|(v_i,v_j)\in E(\mathcal{L}),t=1,2,...,n\} \cup \{(v_i^{t},v_i^{t+1})|v_i\in V(\mathcal{L^*}),t=1,2,...,n-1\}$, where $\mathcal{L^*}$ is a 2-D lattice and $n$ is a positive integer. In particular, we say two edges $e_1=(v_1^{t_1}, v_2^{t_2})$, $e_2=(v_3^{t_3}, v_4^{t_4})$ are parallel in a 3-D lattice if $v_1=v_3$, $v_2=v_4$ and $t_3-t_1=t_4-t_2$, notated by $e_1//e_2$.

Lastly, it is also required to define the path $s$ in a lattice or graph $\mathcal{G}$ to be a sequence of vertices $s=[v_1^{t_1},v_2^{t_2},...,v_n^{t_n}]$, where $(v_k^{t_k},v_{k+1}^{t_{k+1}})\in E(G)$ for any $k\in{1,2,...,n-1}$. The reverse of $s$ is defined as $\Bar{s}=[v_n^{t_n},v_{n-1}^{t_{n-1}},...,v_1^{t_1}]$. We refer to $s^{(k)}$ as the $k$th vertex in path $s$ and refer to the first and last vertices as the endpoints of $s$. If an edge $e$ satisfies $e=(s^{(k)},s^{(k+1)})$, we say edge $e$ is on the path $s$. If the two endpoints of path $s$ are the same, we say $s$ is enclosed. In addition, two functions acting on the paths are defined as follows. First, if $s_1$ and $s_2$ have the same endpoint, we say $s_1$, $s_2$ can be concatenated and $s=s_1+s_2$ is the concatenation of $s_1$ and $s_2$, where $s$ is the sequence of $n+m-1$ vertices satisfying $[s^{(1)},s^{(2)},...,s^{(n)}]=s_1$ or $\Bar{s}_1$, and $[s^{(n)},s^{(n+1)},...,s^{(n+m-1)}]=s_2$ or $\Bar{s}_2$. Second, suppose $s=[v_1^{t_1},v_2^{t_2},...,v_n^{t_n}]$ is a path in a 3-D lattice $\mathcal{L^*}\times n$, the projection of $s$ is defined as $[v_1,v_2,...,v_n]$, denoted by ${\rm proj}(s)$, which is a path in the 2-D lattice $\mathcal{L^*}$. Here the projection is modulo 2, which means that if an even number of edges on $s$ are projected into the same edge, it is equivalent to deleting this edge in the projection.

\begin{figure*}[htb]
\centering
\includegraphics[width=17.5cm]{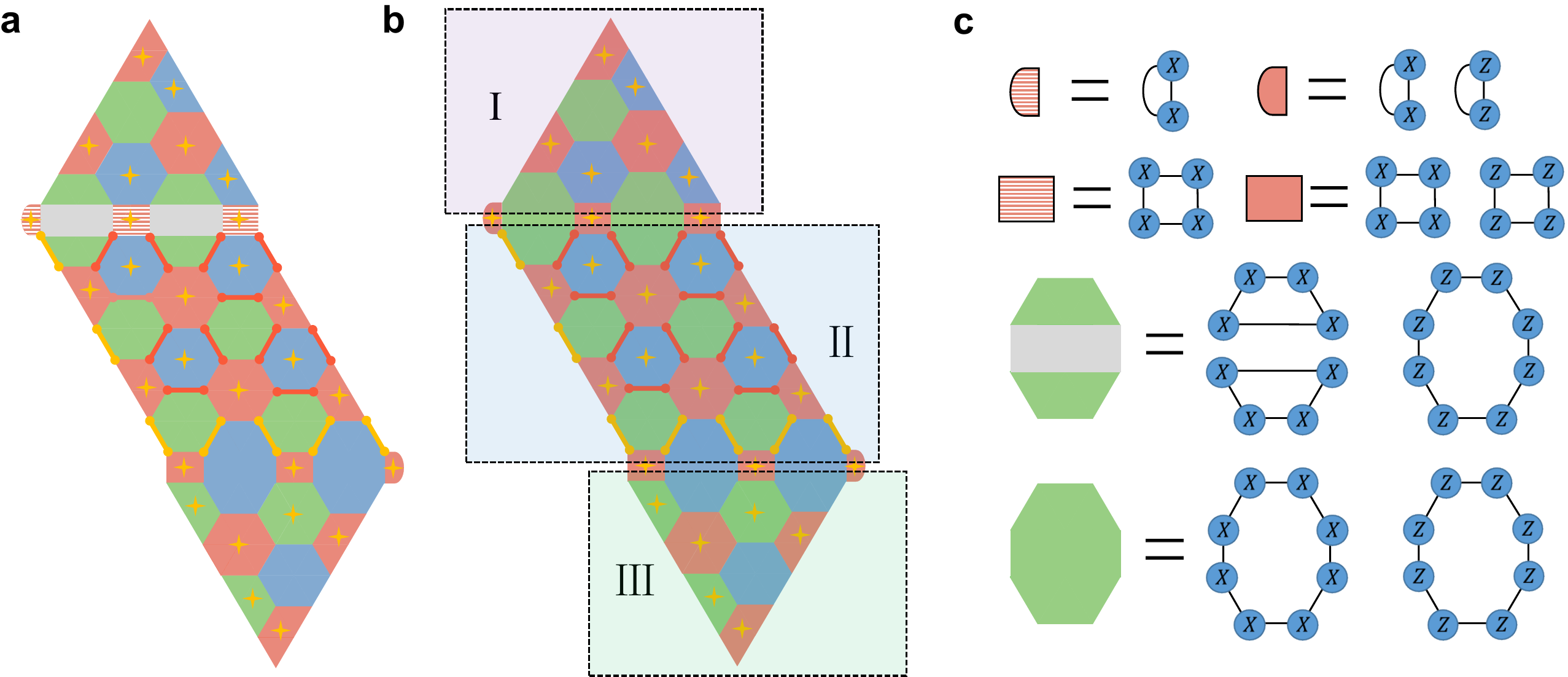} 
\caption{\justify (a) Lattice surgery of triangular color codes for measuring $X_L\otimes X_L$.  The qubits in the middle part are initialized to Bell states (red and yellow pairs). Then the result is the product of the measurement outcomes of starred $X$-type stabilizers. (b) Lattice surgery of triangular color codes for measuring $X_L\otimes X_L$ and $Z_L\otimes Z_L $ in parallel. The results are the product of the measurement outcomes of starred $X$-type (or $Z$-type)  stabilizers. The lattice is divided into three regions, comprising the upper and lower logical qubit regions and the middle auxiliary lattice region (marked by three different colors). At the boundaries between two regions, there exist stabilizer generators with weights of 8 and 4 (or 2).  (c) Specific forms of the stabilizer generators at the boundaries between regions.}\label{fig2}
\end{figure*}

\subsection{Color code lattice surgery and Pauli-based computation}\label{color code lattice surgery and Pauli-based computation}

Lattice surgery \cite{horsman2012surface,landahl2014quantum,thomsen2022low,litinski2018lattice,litinski2019game} is a measurement-based scheme allowing for efficient implementation of universal gate sets by fault-tolerant logical multi-body Pauli operator measurements. It is especially suitable for 2-D topological quantum codes, such as color codes or surface codes, as it only requires 2-D qubit layout and local interactions. 

For triangular color codes, Fig.~\ref{fig2}a shows an example of lattice surgery for measuring $X_{L}\otimes X_{L}$ of two logical qubits. Two triangular logical qubits are merged by an ancillary lattice in the middle, where data qubits are initialized to Bell pairs. Then the QEC cycles are performed to measure the stabilizers given in Fig.~\ref{fig2}c. The outcome of the $X_{L}\otimes X_{L}$ measurement is determined by the product of some stabilizer measurement outcomes, based on $X_{L}\otimes X_{L}=S^*B^*$, where $S^*$ is the product of $X$-type stabilizers labeled by a star and $B^*$ is the product of the stabilizers $X\otimes X$ of yellow Bell states in Fig.~\ref{fig2}a. After repeating $d$ rounds of QEC circuits for fault-tolerance, two logical qubits are split by measuring their own stabilizers. Based on $X_{L}\otimes X_{L}$ measurement and transversal logical Clifford gates of color codes, one can realize arbitrary 2-qubit logical Pauli operator measurements by lattice surgery.

Different from surface codes, lattice surgery of color codes allows arbitrary pairs
of commuting logical Pauli measurements in parallel while the space cost remains the same \cite{thomsen2022low}.
For example, keeping the qubit layout in Fig.~\ref{fig2}a, parallel measurements of $X_{L}\otimes X_{L}$ and $Z_{L}\otimes Z_{L}$ can be achieved as long as the stabilizers that connect logical qubits and ancillary lattice are replaced with the stabilizers shown in Fig.~\ref{fig2}b. Ref.~\cite{thomsen2022low} and Ref.~\cite{kesselring2024anyon} give more general examples of color code lattice surgery with different types of weight-4 and weight-8 stabilizers at the boundaries between regions.

Through lattice surgery, one can perform gate-based computation where a universal gate
set is realized by Pauli operator measurements and ancilla qubits \cite{fowler2018low}. However, with the capacity of measuring arbitrary Pauli operators, a natural way to achieve universal quantum computing is Pauli based computation (PBC) \cite{bravyi2016trading,prabhu2022new,chamberland2022universal}. 

\begin{figure}[htb]
\centering
\includegraphics[width=8cm]{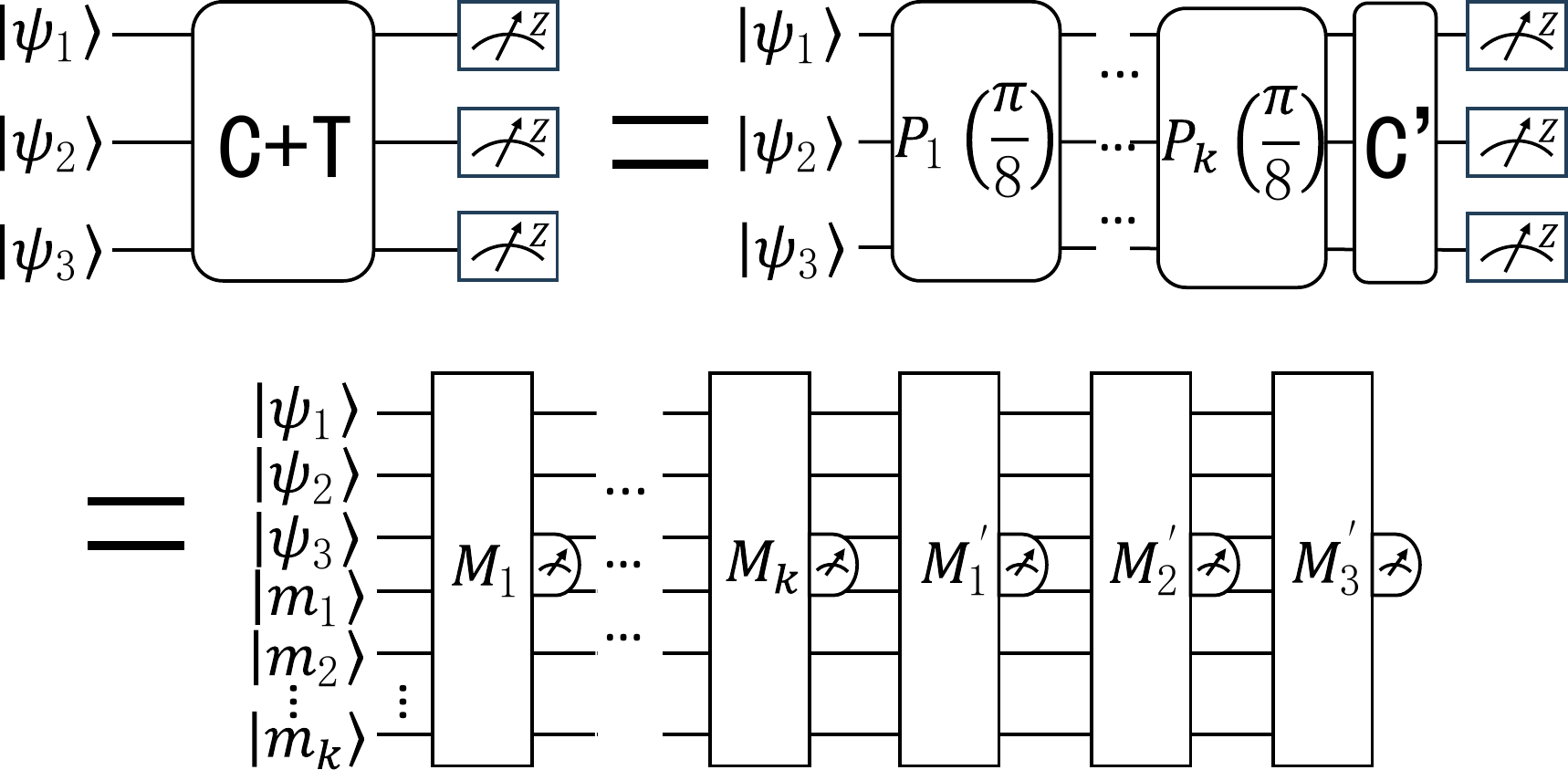}
\caption{\justify Equivalence of the gate-based quantum computation model and the PBC model. Any computation can be expressed as a Clifford+$T$ circuit. First we commute all the Clifford gates to the right of the circuit and then perform all $P(\frac{\pi}{8})$ gates by introducing magic states $\ket{m_i}$, and multi-body Pauli operator measurements $M_i$. Lastly, the Clifford gates can be absorbed by the final measurements.}\label{fig3}
\end{figure}

\begin{figure}[htb]
\centering
\includegraphics[width=8cm]{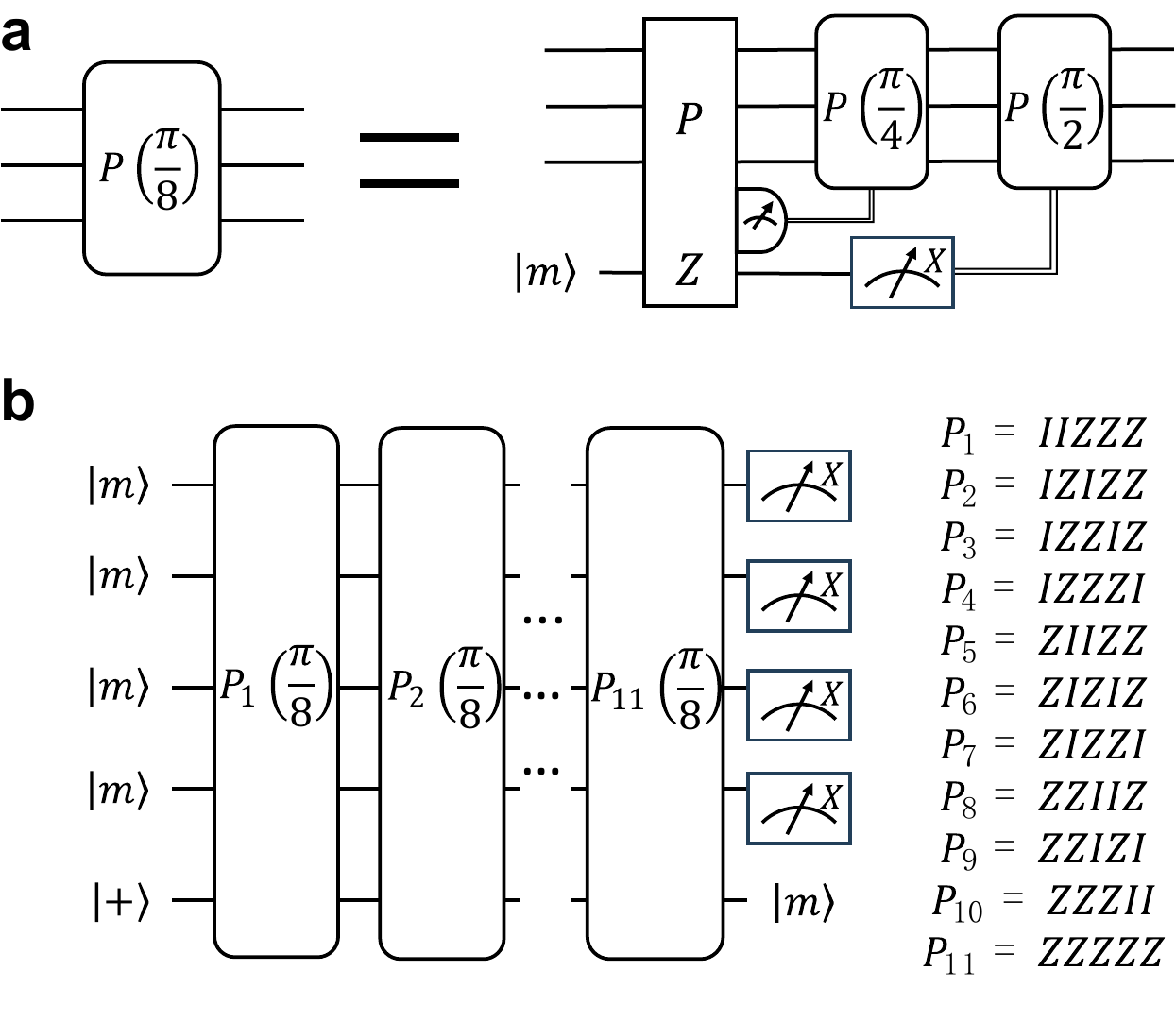}
\caption{\justify  (a) $\pi/8$ rotation gate implemented by consuming a magic state. (b) Circuit of the 15 to 1 magic state distillation protocol. Four magic states are consumed in the initial states and 11 magic states are consumed to implement $P(\frac{\pi}{8})$ gate (total of 15).}\label{fig4}
\end{figure}

In the PBC model, a gate-based quantum circuit is equivalent to a series of Pauli operator measurements on initial states $\ket{0^{\otimes n}}$ and ancilla magic states $\ket{m^{\otimes k}}$. To illustrate this, let us start with a Clifford+$T$ circuit in Fig.~\ref{fig3}, which is well-known to form a universal gate set. By commuting all $T$ gates to the foreside of the circuit, the Clifford+$T$ circuit is replaced with a series of $\frac{\pi}{8}$ Pauli rotations $P(\frac{\pi}{8})=e^{-i\pi P/8}$, followed by Clifford gates $C'$ and $Z$-basis measurements. Then a $\pi/8$ Pauli rotation can be realized by Pauli operator measurement $M_i$ and Clifford gates with an ancilla magic state (see Fig.~\ref{fig4}a). Lastly, we commute all Clifford gate and $Z$-basis measurements, which transforms $Z$-basis measurements to multi-body Pauli
operator measurements. Therefore, we have proven the PBC model is equivalent to a universal gate-base computation model.

Note that the outcome of measurement $M_i$ may affect subsequent measurements, since there is a Clifford gate $P(\frac{\pi}{4})$ controlled by measurement outcome when we perform $P(\frac{\pi}{8})$ gate. Hence, the execution time of the PBC circuit is linearly related to the layers of $P(\frac{\pi}{8})$ gates where $P(\frac{\pi}{8})$ gates commute in the same layer. Some literatures focus on reducing the time overhead. For instance, the time-optimal scheme \cite{fowler2012time,litinski2019game} allows parallelism of the measurement layers by using a large number of ancilla qubits, and the temporally encoded lattice surgery scheme \cite{chamberland2022universal} effectively reduces the time of per measurement.

\subsection{Magic state distillation}\label{magic state distillation}
In general, a practical quantum computer requires a universal gate set. According to the Eastin-Knill theorem \cite{eastin2009restrictions}, no QEC code admits a universal and transversal logical gate set. The logical Clifford gates in the topological quantum codes are typically easier to implement by lattice surgery or braiding operations \cite{fowler2012towards,raussendorf2006fault}. However, performing non-Clifford logical gates such as a $\pi/8$ Pauli rotation gate $P(\frac{\pi}{8})$ requires extra resource state $\ket{m}=\ket{0}+e^{i\pi/4}\ket{1} $ called magic state. 

Utilizing a magic state as the ancilla qubit, one can realize a logical gate $P(\frac{\pi}{8})$ by the circuits in Fig.~\ref{fig4}a using multi-body Pauli operator measurements and classically controlled Clifford gates. Suppose the Pauli operator measurements and Clifford gates are
implemented fault-tolerantly by lattice surgery, the fidelity of the logical gate $P(\frac{\pi}{8})$ is mainly determined by the fidelity of $\ket{m} $. High-fidelity magic states can be produced by several copies of noisier magic states, which is called magic state distillation. Magic state distillation is often considered the most costly part of fault-tolerant quantum computing, even though there is an increasing number of efficient
protocols proposed \cite{jochym2016stacked,brown2020fault,chamberland2019fault,litinski2019magic,chamberland2020very}.

Here, we briefly review the most well-known magic state distillation protocol, the 15 to 1 protocol as an example \cite{litinski2019game}. This protocol is based on the 15-qubit Reed-Muller code, the smallest QEC code with the transversal $T$ gate \cite{koutsioumpas2022smallest}. The distillation circuit starts with four magic states $\ket{m}$ and a $\ket{+}$ state, followed by $\pi/8$ Pauli rotation gates and $X$-basis measurements, as shown in Fig.~\ref{fig4}b. Since one magic state is consumed in each $\pi/8$ Pauli rotation gate implementation, a total of 15 magic states are consumed to distill one high-fidelity magic state. The errors of at most two of 15 magic states can be detected by the $X$-basis measurements. If any $X$-basis measurement outcome is $-1$, all qubits are discarded and the distillation protocol is restarted. Suppose that every input magic state suffers Pauli $Z$ error with probability $p_i$, the error rate of the output magic state is $p_{out}\approx 35p_i^3$ since there are 35 combinations of three faulty magic states that cannot be detected. If the input magic states suffer $X$, $Y$, $Z$ Pauli errors with probabilities $p_x$, $p_y$, $p_z$ respectively, the output error rate is $p_{out}\approx 35[p_z+\frac{( p_x+ p_y)}{2}]^3$ \cite{litinski2019magic}.

Typically, the initial input magic states are produced by a non-fault-tolerant procedure called state injection. This procedure injects an arbitrary physical state $\ket{\psi}=\alpha\ket{0}+\beta\ket{1}$ into the logical state $\ket{\psi_L}=\alpha\ket{0_L}+\beta\ket{1_L}$. The state injection is crucial since the fidelity of the distilled magic states depends strongly on the quality of the initial magic states.  For example, in the 15 to 1 distillation protocol, if the error rate of the initial states $p_i$ is reduced to $ p_i /n$, the error rate of the output state is reduced by around $n^{(3^{k})}$ times for $k$ rounds of distillation.

\subsection{Qubit layout and circuit-level noise model}\label{circuit-level noise model}
Here we introduce the basic assumptions about the qubit layout. The data qubit is placed in each vertex of the primal lattice of triangular color code, and two syndrome qubits are in each colored face for the stabilizer measurements. One of the syndrome qubits is initialized to $\ket{0}$ for measuring the $Z$-type stabilizer and the other is initialized to $\ket{+}$ for measuring the $X$-type stabilizer. The syndrome qubits are coupled with data qubits in the corresponding faces by CNOT gates. Those assumptions run through the following discussions of the color code decoding, lattice surgery and magic state injection. 

\begin{figure}[b]
\centering
\includegraphics[width=8cm]{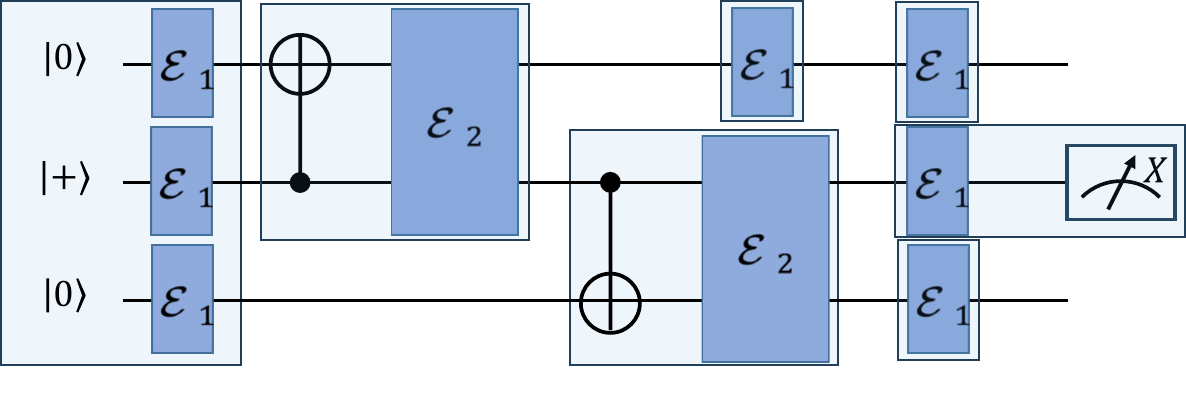}
\caption{\justify An example of the noisy circuit to prepare a Bell state. We label the noise channels of the initialization errors, two-qubit gate errors, idling errors and measurement errors separately.}\label{fig5}
\end{figure}

It is known that error correcting performance of QEC codes depends largely on noise models. Throughout this paper, the noise model we considered is the circuit-level depolarizing noise model. The depolarizing error channels are defined as:
\begin{equation}
\begin{aligned}
\mathcal{E}_{1}(\rho_1)&=(1-p)\rho_1+(p/3)\sum_{P\in\{X,Y,Z\}}P\rho_1P\\
\mathcal{E}_{2}(\rho_2)&=(1-p)\rho_2+(p/15)\\
&\times \sum_{\substack{P_1,P_2\in\{I,X,Y,Z\},\\P_1\otimes P_2\neq I \otimes I}}P_1\otimes P_2\rho_2P_1\otimes P_2,
\end{aligned}
\end{equation}
where $\rho_1$ and $\rho_2$ are single-qubit and two-qubit density matrices respectively and $p$ is the physical error rate.

When simulating a noisy quantum circuit, we approximate every noisy operation by an ideal operation followed by the depolarizing error channel $\mathcal{E}_{1}$ or $\mathcal{E}_{2}$. Specifically, the depolarizing error channels are added to the circuits by the following rules: 

(1) add $\mathcal{E}_{1}$ after preparing each $\ket{0}$ or $\ket{+}$ state;

(2) add $\mathcal{E}_{1}$ after each idle operation;

(3) add $\mathcal{E}_{1}$ before each measurement in $Z$- or $X$-basis;

(4) add $\mathcal{E}_{2}$ after each CNOT gate.

A noisy quantum circuit for preparing a bell state is shown in Fig.$\,$\ref{fig5}. Note that if a qubit in some step is not acted by any of the preparation, two-qubit gate or measurement while operations of other qubits are being performed, it is assumed to be applied to an idle operation and suffer the error channel $\mathcal{E}_{1}$. In fact, errors from idle operations account for a large portion when we simulate  high-weight stabilizer measurement circuit of the color code.

\section{Improved Color code decoding}\label{Color code decoding}
This section describes the improved color code decoding strategy and presents numerical results. Our improved decoding strategy is based on the projection decoder \cite{beverland2021cost,delfosse2014decoding}. The variation is that, we execute the MWPM algorithm on the graphs with error-rate-related weights (referred to as decoding graphs below), rather than the original dual lattice $\mathcal{L^*} \times n$. In Sec~\ref{CNOT schedule}, we discuss how to construct the decoding graphs and the criterion that a proper CNOT schedule (i.e., the order of the CNOT gates in the QEC circuits) needs to meet. In Sec~\ref{decoding algorithm}, we explain the entire process of the decoding algorithm and give the numerical results.

\subsection{Decoding graphs and CNOT schedules}\label{CNOT schedule}

\begin{figure}[t]
\centering
\includegraphics[width=8cm]{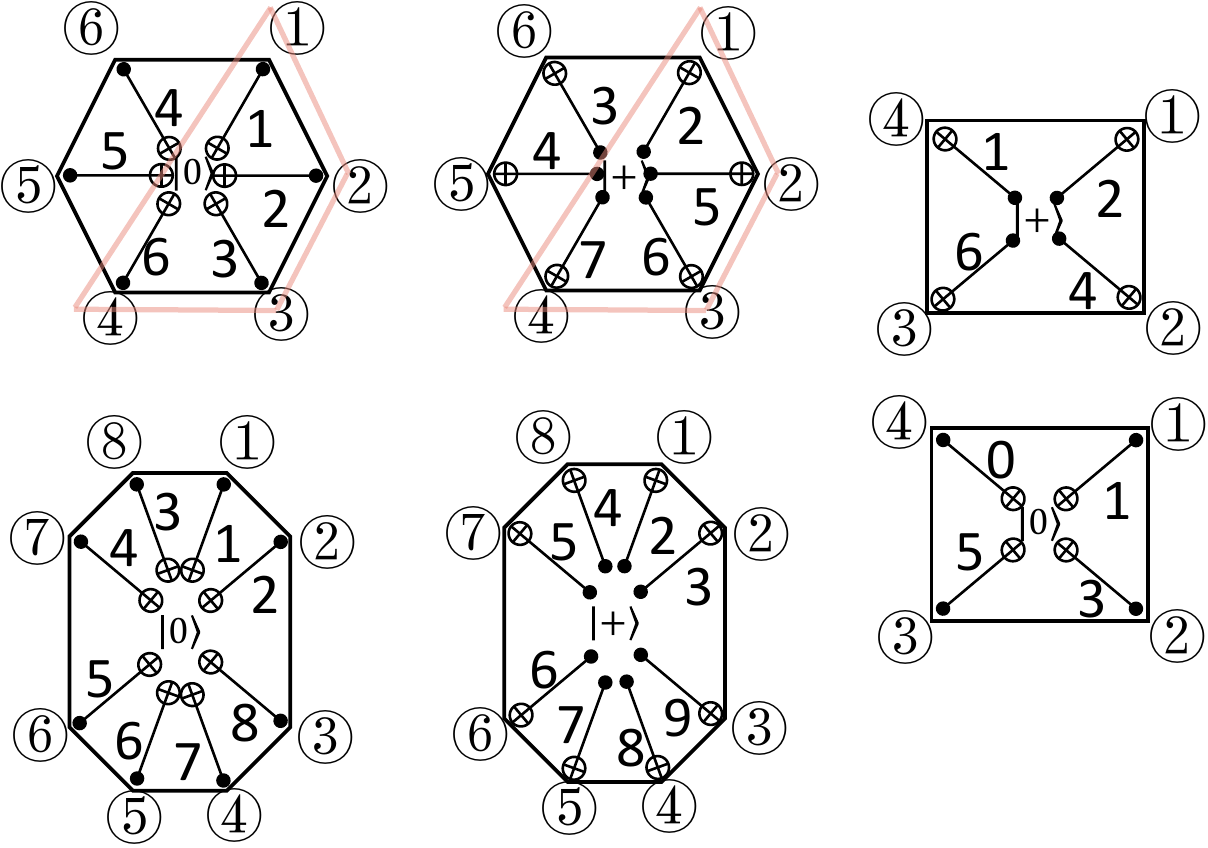}
\caption{\justify CNOT schedules of stabilizers with different weights used in color code QEC circuits and lattice surgery. The outside and inside numbers indicate position labels and time orders, respectively. The CNOT schedules for weight-4 stabilizers on the triangular boundary can be derived by partitioning the CNOT schedules for weight-6 stabilizers. For example, the red trapezoid in the figure corresponds to the weight-4 stabilizer on the left boundary of the triangle. Similarly, the CNOT schedules of the stabilizers on the right and bottom boundaries of the triangle can be derived by partitioning the hexagon into trapezoids in different orientations.}\label{fig6}
\end{figure}

\begin{figure}[t]
\centering
\includegraphics[width=8cm]{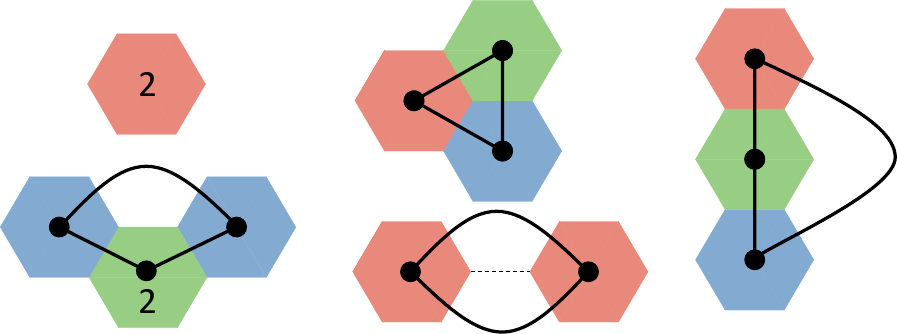}
\caption{\justify Five possible syndrome changes (colored faces) and their connections (solid lines) of a single error in the color code QEC circuit with the proper CNOT schedule. We only show the projection of the faces, which means that the syndrome changes may come from different time layers. A face labeled 2 indicates that the syndrome changes twice in two consecutive time layers and corresponds to a vertical connection.}\label{fig7}
\end{figure}

When decoding topological QEC codes by the MWPM algorithm, the performance of the decoder is affected by the matching weights. For example, in the surface code, the MWPM algorithm is typically applied in a graph with weights where the weight of the edge is related to the probability of the given pair of syndrome changes \cite{wang2011surface,fowler2012topological,fowler2012towards,higgott2023improved,criger2018multi}. Ref.~\cite{wang2011surface} has calculated the probabilities of the six types of edges by categorically analyzing the error in the surface code QEC circuit. 

By introducing the error-rate-related weights, the decoding performance increases substantially. Therefore, this inspires us to consider introducing the error-rate-related weights in the color code decoding graph. However, constructing such decoding graphs of color codes faces two main challenges:

(a) The QEC circuit of the color code is deeper and the qubit array is more complex compared to that of surface codes, which makes analyzing the edge error rate categorically more difficult.

(b) For the surface code, by setting the proper CNOT schedule, a single error in the circuit always causes two syndrome changes or less. Here a single error means the Pauli
error from one depolarizing error channel in the noisy circuit. However, for the color code, because of the high-weight stabilizer measurements and error propagations, a single error may cause syndrome changes of more than two, in which case an error no longer corresponds to just one edge in the decoding graph.

To address challenge (a), we designed an automated procedure to count the syndromes of various errors in the QEC circuit. First, we construct three 3-D decoding graphs $\mathcal{G}_{C}$ ($C\in \{RG,RB,GB\}$), where the vertical dimension is the time layers, and in each time layer is a 2-D graph. We assume that the total number of the time layers in the decoding graph $\mathcal{G}_{C}$ is $d+1$, which corresponds to $d$ rounds of noisy QEC circuits and one round of perfect QEC circuit. The vertices of the decoding graph are the same as those in the 3-D restricted lattice $\mathcal{L}^*_{C}\times(d+1)$, but the edges are different. 

Second, the edges in the decoding graph $\mathcal{G}_{C}$ are determined by the single error in the QEC circuit. To establish the relation between a single error and an edge, we simulated a perfect circuit attached to a single error $\epsilon$ in the different positions in the QEC circuit. This single error could originate from channel $\mathcal{E}_{1}$ ($\epsilon\in\{X, Y,Z\}$) or from channel $\mathcal{E}_{2}$ ($\epsilon\in\{I,X,Y,Z\}^{\otimes2}/I^{\otimes2}$). For each possible position and type of the single error, the procedure simulates the QEC circuit one time and records the syndrome changes in each simulation.

Suppose that in the decoding graph $\mathcal{G}_{C}$, these recorded vertices are connected in pairs by some rules (see Appendix~\ref{ab} for details), and the error $\epsilon$ leads to the edge $e$. Then the weight of $e$ is assigned $w_{e}=-\log \sum_{\epsilon} p_\epsilon(e)$, where $p_\epsilon(e)=p/3$ if $\epsilon$ is a single-qubit error, otherwise $p_\epsilon(e)=p/15$.

Now let us consider challenge (b): how to reduce the number of syndrome changes caused by a single error. In fact, the problem can be dealt with by a proper CNOT schedule. Intuitively, we hope that the multiple errors resulting from a single error propagation are located in neighbor vertices on a face of the primary lattice, so that some syndrome changes can cancel each other out. For example, let us consider the measurement circuit of the stabilizer generator on an hexagonal face. If an $X$ error in syndrome qubit propagates to two neighbor data qubits, the number of syndrome changes is two. In contrast, if two error data qubits are not neighbors, the number of syndrome changes is four. Based on this intuition, we propose a criterion that a proper CNOT schedule needs to meet. 

Suppose $l_i$ is the position label of the data qubit where the $i$th CNOT gate ($i=1,2,3,\ldots$) is applied, with ${l_i} = {1, 2, \ldots, N}$ representing integers fixed on the weight-$N$ stabilizer that increase clockwise (see Fig.~\ref{fig6}). A proper CNOT schedule is defined as: given any $i\in\{1,2,…,N\}$, $\forall a,b\in\{l_{i+1},l_{i+2},…,l_N\}$, we have $|a-b|\,{\rm mod}\, N\leq N-i$, where $x\,{\rm mod}\, N$ is the modulo function ranged in $[0,N)$. In other words, the criterion guarantees that the syndrome qubit’s error after the $i$th CNOT gate always propagates to the data qubits whose positions $\{l_{i+1},l_{i+2},…,l_N\}$ are $N-i$ neighbor positions. It can be verified that when $N$ takes values of 2, 4, 6, or 8 (corresponding to all the stabilizer generators involved in this paper), the syndrome changes of these $N-i$ errors on each 2-D restricted lattice do not exceed 2.

Previous work \cite{beverland2021cost} identified the optimal CNOT schedule with a specific physical error rate and qubit layout through numerical testing, which exactly meets the criterion. Therefore, we continue to use this CNOT schedule of weight-6 stabilizers in Fig.~\ref{fig6}. It should be noted that the criterion we proposed also applies to higher-weight stabilizer measurements, such as the weight-8 stabilizers used in the lattice surgery (see Sec \ref{Lattice surgery} and Fig.~\ref{fig6}).

Once the CNOT schedule is given, we can explain how to connect the syndrome changes in pairs of a single error. There are 5 types of syndrome changes of a single error using the well-selected CNOT schedule. As explained in Appendix~\ref{ab} and Fig~\ref{fig7}, the edges $e\in E(\mathcal{G}_{C})$ in the three decoding graphs are determined by the circuit error that connects the syndrome change pairs. Except for one type in the bottom left corner of Fig~\ref{fig7}, only two syndrome changes occur in each decoding graph at most. Moreover, if the error propagates to the boundary data qubits, one of the vertices of the edge may come from the boundary vertex set $\{v_R^{(t)}, v_G^{(t)}, v_B^{(t)}\}$, where $t$ is the time layer in which the error occurs.

\subsection{Improved color code decoding algorithm}\label{decoding algorithm}
In this section, we show the whole decoding process of triangular color codes. Without loss of generality, we assume that one corrects $X$-type errors using the $Z$-type syndromes. Roughly speaking, the decoding algorithm can be divided into three parts. First, we construct three decoding graphs and apply the MWPM algorithm on them. Then the matching results will be mapped to the 3-D lattice $\mathcal{L}^*_{C}\times(d+1)$. Finally, the error data qubits are determined by combining the matching results.

The inputs of the decoding algorithm are code distance $d$, physical error rate $p$ and syndrome changes $\sigma \subseteq V[\mathcal{L}^*\times(d+1)]$. In order to get the error correction set $R\subseteq F(\mathcal{L}^*)$, the decoding algorithm is performed in the following steps.

\begin{itemize}
    \item [1.]
    Input $d$, $p$ and syndrome changes $\sigma$, and construct the 2-D dual lattice $\mathcal{L}^*$. For $C\in \{RG,RB,GB\}$, construct the 3-D dual lattice $\mathcal{L}^*\times(d+1)$, the 3-D restricted lattice $\mathcal{L}^*_{C}\times(d+1)$ and decoding graph $\mathcal{G}_{C}$.
    \item [2.]
    For $C\in \{RG,RB,GB\}$, apply the MWPM algorithm on $\mathcal{G}_{C}$ to pair up the vertices in $\sigma_C=\sigma\cap V(\mathcal{G}_{C})$ and obtain a path set $S_C$.

    \item [3.]
    For all $s$ in $S_C$, check whether the edge $e\in E[\mathcal{G}_{C}]$  satisfies $e\in E[\mathcal{L}^*_{C}\times(d+1)]$. If not, replace $e$ on $s$ with the shortest path connecting endpoints of $e$ in $\mathcal{L}^*_{C}\times(d+1)$.
    
    \item [4.]
    Combine path sets $S=S_{RG}\cup S_{RB}\cup S_{GB}$. For $s_1$, $s_2$ in $S$, if $s_1,s_2$ can be concatenated, replace $s_1$, $s_2$ of $S$ with their concatenation $s_1+s_2$. Repeat this step until all the paths in $S$ cannot be concatenated.
    \item [5.]
    For path $s\in S$, project $s$ to the 2-D lattice $\mathcal{L}^*$. The projection ${\rm proj}(s)$ will divide $\mathcal{L}^*$ into two face sets $F_1$ and $F[\mathcal{L}^*]\backslash F_1$, and select the smaller one as the correction set $R_s$. Output the total correction set $R=\oplus_s R_s$, where $\oplus$ is modulo 2 addition. 
\end{itemize}

\begin{figure}[tb]
\centering
\includegraphics[width=8cm]{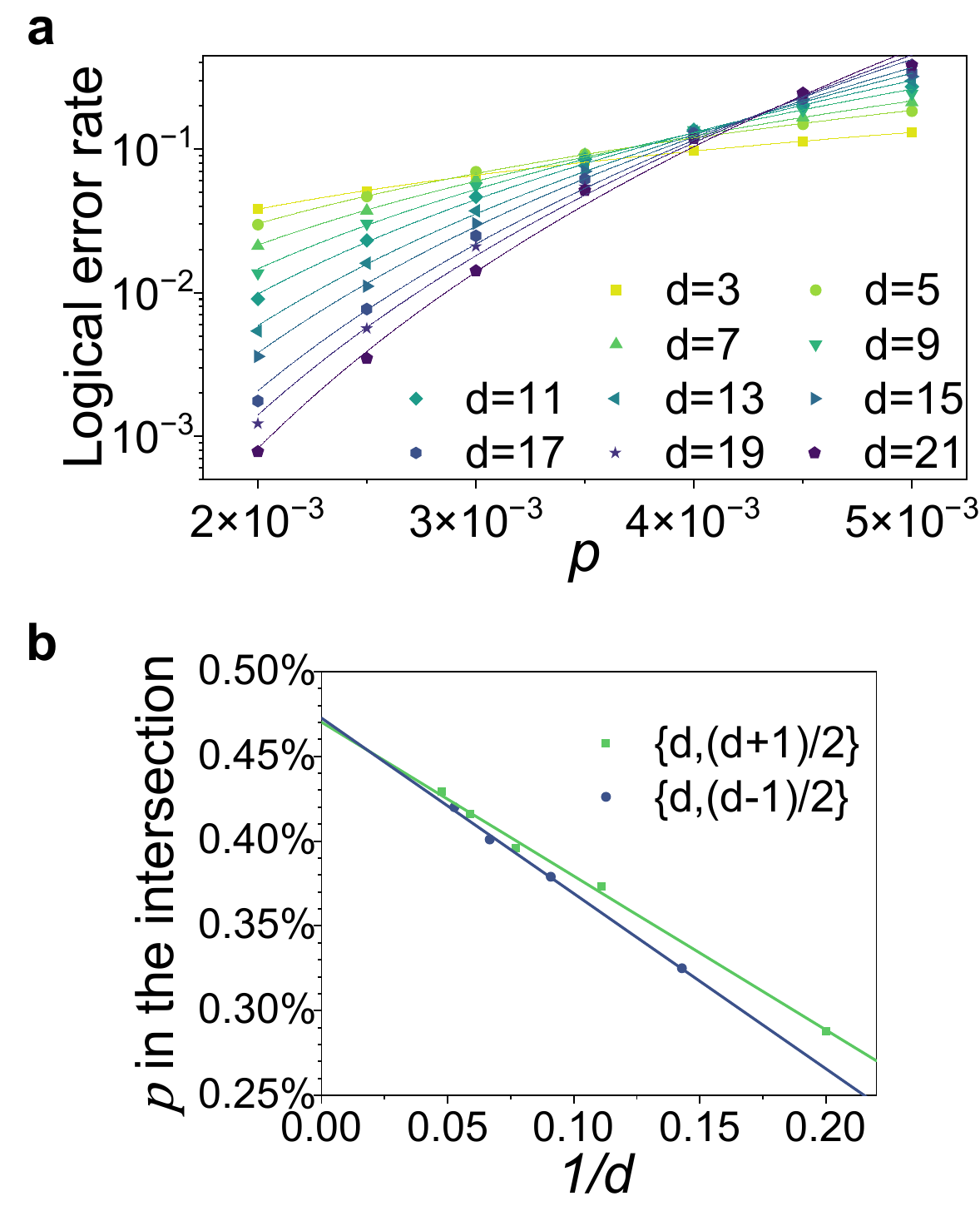}
\caption{\justify (a) Logical error rates $P_L$ of the triangular color code under the circuit-level noise model for various code distances $d$. The curves are fitted by the ansatz $P_L=\alpha p^{\beta}$, where $\alpha$ and $\beta$ vary with $d$. 
(b) The intersections of pairs of curves of $P_L$ with distances $\{d,(d+1)/2\}$ and $\{d,(d-1)/2\}$. From the linear extrapolation of the data, the threshold is around $0.47\%$.}\label{fig8}
\end{figure}

The algorithm essentially follows the steps outlined in Ref.~\cite{beverland2021cost}. The distinction is that we introduce decoding graphs with error-rate-related weights and map it back to a 3-D lattice $\mathcal{L}^*_{C}\times(d+1)$ in the 3rd step.

After the 4th step, the paths in $S$ are either enclosed or have two endpoints from the boundary vertex set $\Tilde{V}=\{v_x^t|v_x\in\{v_R,v_B,v_G\},t=1,2,...,d+1\}$. Therefore, after the projection in the 5th step, the path always divides $\mathcal{L}^*$ into two complementary regions.

We also remark that the entire algorithm can be executed efficiently in polynomial time. Firstly, when constructing the decoding graph $\mathcal{G}_{C}$, it is necessary to simulate approximately $\mathcal{O}(m)$ QEC circuits, where $m$ is the number of data or syndrome qubits. This simulation frequency is proportional to the number of noise channels in the QEC circuit and consequently, proportional to the number of qubits. The time complexity for simulating one QEC circuit is $\mathcal{O}(m^2)$ \cite{aaronson2004improved}, resulting in an overall time complexity of $\mathcal{O}(m^3)$. For given $d$ and $p$, the decoding graphs only need to be generated one time and can be repeatedly used. Additionally, the primary time consumption in the decoding algorithm occurs in the MWPM algorithm, which has a time complexity of $\mathcal{O}(m^3\log m)$ \cite{higgott2022pymatching}. Overall, the time complexity of the decoding algorithm is $\mathcal{O}(m^3\log m)$.

We simulate the logical error rates $P_L$ of the color code with code distance from $d=3$ to $d=21$ and fit the curves by ansatz $P_L=\alpha p^\beta$, where parameters $\alpha$ and $\beta$ vary with different $d$ (see Fig.~\ref{fig8}a). Theoretically, the threshold is the value of the physical error rate $p$ of the intersection of the curves when $d\rightarrow\infty$. Here the threshold is estimated progressively by the intersections of two classes of the logical error rate curves, similar to the method in Ref.~\cite{beverland2021cost}. We select the curves with distances $\{d,(d+1)/2\}$ and $\{d,(d-1)/2\}$, since their slopes differ quite remarkably to identify the intersections (see Fig.~\ref{fig8}b). From a linear extrapolation of the data, the threshold is around $0.47\%$. For the convenience of future research, we provide a formula for the average logical error rate per round of QEC circuit, $\Bar{P_L}=0.018({p}/{0.47\%})^{d/3+0.04}$ , obtained through fitting data with $d\geq 11$. 

Intuitively, our strategy improves upon the decoder presented in Ref.~\cite{beverland2021cost}. However, we emphasize that the threshold result here cannot be directly compared to the 0.37\% threshold in Ref.~\cite{beverland2021cost} because our noise model is slightly weaker. Nonetheless, we provide some simulation results under the same noise model in Appendix~\ref{dc} to directly demonstrate the performance improvement.

\begin{figure}[t]
\centering
\includegraphics[width=5cm]{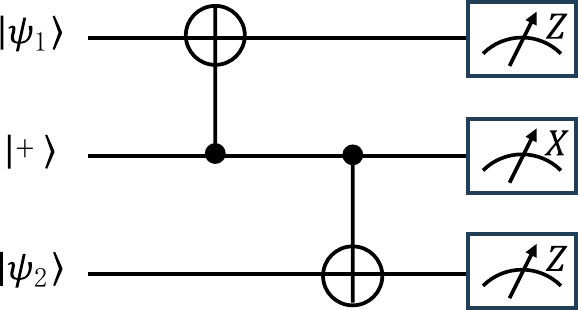}
\caption{\justify Circuit to measure $X\otimes X$ and $Z\otimes Z$ using a $\ket{+}$ state as ancilla. The result of $X\otimes X$ is the measurement outcome on $\ket{+}$, and the result of $Z\otimes Z$ is the product of measurement outcomes on $\ket{\psi_1}$ and $\ket{\psi_2}$. After the measurements, all the states will be discarded. Note that this circuit can be viewed as the inverse of the circuit in Fig~\ref{fig5}.}\label{fig9}
\end{figure}

\section{Circuit-level decoding of color code lattice surgery }\label{Lattice surgery}

\begin{figure*}[htb]
\centering
\includegraphics[width=17cm]{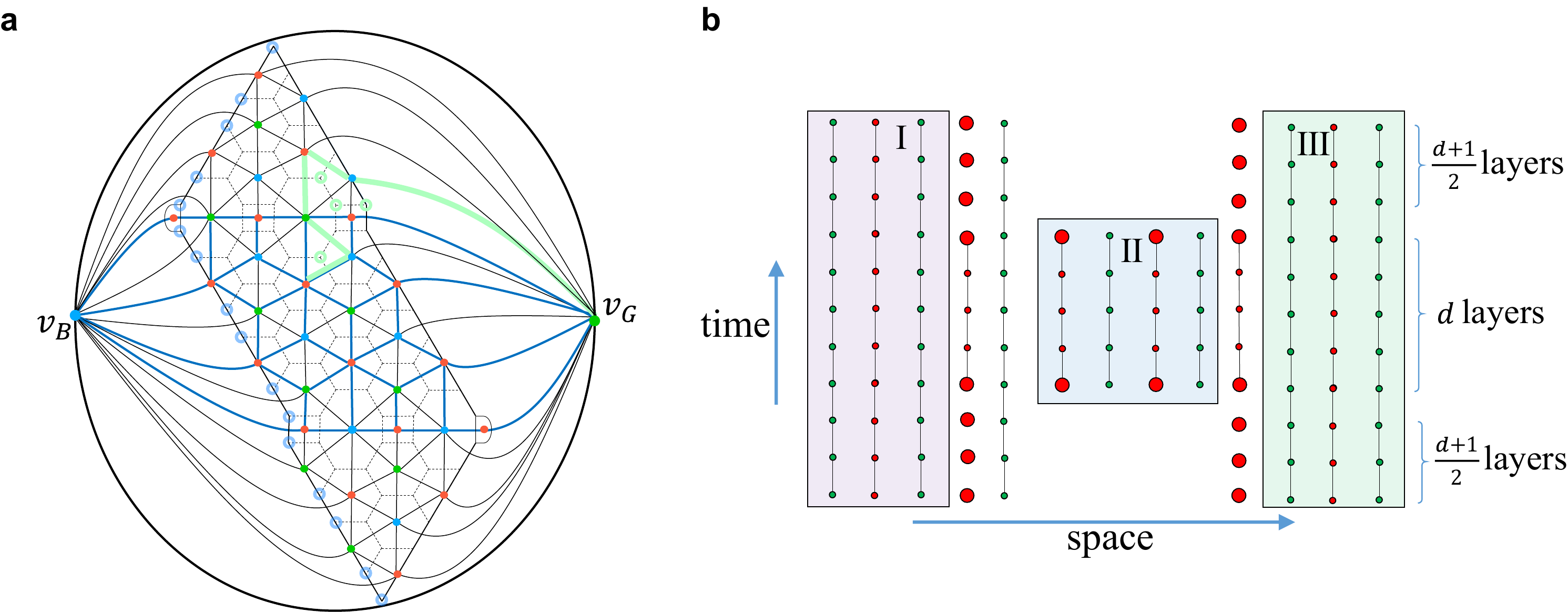}
\caption{\justify (a) The merged dual lattice $\mathcal{L}^{m}$ of lattice surgery of color code lattice surgery for measuring $X_L\otimes X_L$ and $Z_L\otimes Z_L$ in parallel. The lattice is divided into several subareas by the blue lines and the outermost circle.  An example of the path that ends inside the lattice and corresponding correction are illustrated by green lines and circles. Additionally, the blue circles indicate all the data qubits on the left boundary, which support the boundary operator \(b_B\). (b) Vertices in decoding graph $\mathcal{G}^{ls}_{RG}$ (or lattice $\mathcal{L}^{ls}_{RG}$). Two dimensions in space are compressed into one. The boundary vertex $v_G$ is not shown, although it is also in each layer of $\mathcal{G}^{ls}_{RG}$ or $\mathcal{L}^{ls}_{RG}$. In our simulation, regions \Rmnum{1} and \Rmnum{3} are split in the first and last $(d+1)/2$ layers and  merged by  region \Rmnum{2} in the middle $d$ layers. The magnified vertices represent red vertices in the boundary vertex set $\Tilde{V}$. Beyond the three regions, the vertices in $V_R^m\cup V_G^m$ are retained in each layer of the split steps. To avoid clutter, only vertical edges are shown, while horizontal edges as well as diagonal edges in $\mathcal{G}^{ls}_{RG}$ or $\mathcal{L}^{ls}_{RG}$ are omitted.}\label{fig10}
\end{figure*}

Although in principle the lattice surgery of triangular color codes is feasible, it still requires a circuit-level decoding strategy for its practical application. In this section, we focus on the decoding process during the measurement of two-body Pauli operators using lattice surgery. Sec~\ref{Decoding lattice} presents the detailed steps of the decoding algorithm. Sec~\ref{further discussion} provides numerical results and discusses the applicability of the algorithm.

\subsection{Decoding algorithm}\label{Decoding lattice}
Let us clarify the decoding strategy in the specific example of Fig.~\ref{fig2}b, where two-qubit logical operators $X_{L}\otimes X_{L}$ and $Z_{L}\otimes Z_{L}$ are measured in parallel. Suppose the logical qubits are in the PBC model, $d$ rounds of QEC circuits will be implemented alternately in the merged and split lattice. Here, we consider a section in this process with a total of $2d+1$ layers. Namely, the first $(d+1)/2$ layers and the last $(d+1)/2$ layers are split lattices, while the middle $d$ layers are merged lattices. In the QEC circuits, we continue to use CNOT schedules of weight-6 stabilizers, and give weight-8 and weight-4 CNOT schedules in Fig.~\ref{fig6}, which meets the criterion proposed in Sec~\ref{CNOT schedule}. 

At the beginning and end of the merged step, the Bell pairs are prepared and measured by the circuits in Fig.~\ref{fig5} and Fig.~\ref{fig9}. Note that there are enough syndrome qubits with our qubit layout to act as ancilla in these two circuits, since the number of syndrome qubits is approximately $2n_f$ and the number of Bell states is approximately $n_f$, where $n_f$ is the number of the faces in the middle region. Also remarkably, at the beginning and end of the merged step, the syndromes of green and blue stabilizers are inferred by the product of stabilizers of the Bell states after initialization or measurements, making them noiseless. On the other hand, the red syndromes are random during these two steps, meaning they have no determined initial and final values. After measuring Bell state measurements, the weight-8 stabilizers will transform into weight-4 stabilizers in the next QEC round. Therefore, in the first QEC round of the split lattice, the syndromes of the green or blue stabilizers at the boundaries between regions that used to be weight-8 stabilizers in the merged lattice need to be multiplied by the outcomes of two Bell pair measurements.

Before introducing the decoding algorithm, we specify some notations about lattices and vertices. We use $\mathcal{L}^{m}$ ($\mathcal{L}^{s}$), $\mathcal{L}^{m}_{C}$ ($\mathcal{L}^{s}_{C}$) and $\mathcal{L}^{ls}_{C}$ to denote the 2-D merged (split) dual lattice, the 2-D restricted merged (split) lattice and the 3-D restricted lattice, respectively, where $C\in \{RG,RB,GB\}$. Note that all the lattices discussed in this section are dual lattices, so we omit the superscript asterisk in lattice labeling.  Additionally, when referring to a 3-D lattice, we assume there are only vertically connected edges in the vertical direction, without diagonal edges (diagonal edges will appear in the decoding graphs). In the 3-D restricted lattice $\mathcal{L}^{ls}_{C}$, the vertices of $d$ layers in the middle are from the merged lattice $\mathcal{L}^{m}_{C}$ and the vertices in other layers are from the split lattice $\mathcal{L}^{s}_{C}$ (see Fig.~\ref{fig10}). In particular, attention is paid to the red vertices in the lattice $\mathcal{L}^{m}$. The red vertices in region $k$ ($k=1,2,3$) forms a vertex set denoted by $V_R^k$. Beyond the three regions, there are some weight-2, weight-4 or weight-8 stabilizers at the boundaries between regions, corresponding to vertices in set $V_K^m$ ($K\in \{R,G,B\}$ indicating the color).
We also define a boundary vertex set: $\Tilde{V}=\{v_x^t|v_x\in\{v_B,v_G\}\cup V_R^m,\,t=1,2,...,2d+1\}\cup
\{v_x^t|v_x\in V_R^2,t=(d+3)/2\,\, {\rm or} \,\,(3d+3)/2\}$.

There are two goals of the lattice surgery decoding algorithm. One is obtaining the correction set $R\subseteq F(\mathcal{L}^m)$ as accurately as possible and the other is obtaining the correct measurement results of $X_{L}\otimes X_{L}$ and $Z_{L}\otimes Z_{L}$. Overall, our decoding strategy has three main parts. At the beginning, we construct three decoding graphs $\mathcal{G}_{C}$ ($C\in \{RG, RB, GB\}$) with $2d+1$ layers (see Fig.~\ref{fig10}b). The edges and their weights in $\mathcal{G}_{C}$ can be obtained by simulating lattice surgery circuits by the automated procedure in a similar method as it in Sec~\ref{decoding algorithm}. Then we input the syndrome changes $\sigma$ and apply the MWPM algorithm. Likewise, we map the matching results to 3-D lattice $\mathcal{L}^{ls}_{C}$. Lastly, each path is projected on the 2-D merged dual lattice $\mathcal{L}^{m}$. 

Different from Sec~\ref{decoding algorithm}, after the projection, there are three types of paths instead of two. The first is enclosed and the second starts and ends with $v_G$ or $v_B$. These two types of paths can divide the lattice into two complementary regions and we select the smaller one as the correction. The third type is the path that
starts or ends with the vertices in $V_R^m \cup V_R^2$ and not enclosed (see the green line in Fig.~\ref{fig10}a). This results from the indeterminate measurement results in $V_R^m$ and $V_R^2$, as their initial values are random and there is no perfect measurement in the final round. The syndromes on $V_R^m \cup V_R^2$ correspond to $X\otimes X$ (or $Z\otimes Z$) errors on several Bell pairs (see Fig.~\ref{fig2}). Therefore, the incorrect measurements on $V_R^m \cup V_R^2$ is equivalent to errors on several Bell pairs after decoding. However, as described in Appendix~\ref{ad}, the impact of $X\otimes X$ (or $Z\otimes Z$) errors within Bell pairs on the decoding outcome is trivial. Hence, in the decoding, the lattice $\mathcal{L}^{m}$ is divided into several subareas by the blue lines and the outermost circle (see Fig.~\ref{fig10}a). Each subarea corresponds to a Bell pair or a logical qubit and the error suffusing any subareas is trivial or a logical error. We find the smaller regions separated by the path in each subarea as the correction (see green circles in Fig.~\ref{fig10}a).

As mentioned, the measurement outcome of $X_{L}\otimes X_{L}$ (or $Z_{L}\otimes Z_{L}$) is also required to be determined. We use the product of all measurement outcomes of the starred stabilizers in the $(d+3)/2$  layer as the raw result. Apparently, the raw result is inaccurate unless there are no errors affecting the measurement results. Therefore, if any syndrome change of the starred stabilizer occurs before or in the $(d+3)/2$ layer, we flip the raw result. Besides, we specifically consider the third type of paths mentioned above. The syndrome of the correction  $R_s$ obtained from the third type of paths and the actual syndrome change may be different in some boundary vertices from $V_R^m \cup V_R^2$. If such
vertices are before or in the $(d+3)/2$ layer and the number of these vertices is odd, the raw result also needs to be flipped.

In summary, the decoding algorithm is applied as follows.

\begin{itemize}
    \item [1.]
    Input $d$, $p$ and syndrome changes $\sigma$ and construct the 2-D merged dual lattice $\mathcal{L}^{m}$. For $C\in \{RG,RB,GB\}$, construct the 3-D restricted lattices $\mathcal{L}^{ls}_{C}$ and decoding graph $\mathcal{G}^{ls}_{C}$ by the automated procedure in the similar way as before.
    \item [2.]
    For $C\in \{RG,RB,GB\}$, apply the MWPM algorithm on $\mathcal{G}^{ls}_{C}$ to pair up the vertices in $\sigma_C=\sigma\cap V(\mathcal{G}^{ls}_{C})$ and obtain a path set $S_{C}$.
    \item [3.]
    For path $s\in S_C$, check whether the edge $e\in E(\mathcal{G}^{ls}_{C})$ on $s$ satisfies $e\in E(\mathcal{L}^{ls}_{C})$. If not, replace $e$ with the shortest path connecting endpoints of $e$ in $\mathcal{L}^{ls}_{C}$.
    \item [4.]
    Combine path sets $S=S_{RG}\cup S_{RB}\cup S_{GB}$. For $s_1$, $s_2$ in $S$, if $s_1,s_2$ can be concatenated, replace $s_1$, $s_2$ of $S$ with their concatenation $s_1+s_2$. Repeat this operation until all the paths in $S$ cannot be concatenated.
    \item [5.]
    For path $s\in S$, projects $s$ to $\mathcal{L}^{m}$. If ${\rm proj}(s)$ divide $\mathcal{L}^{m}$ into two regions, select the smaller one as the correction. Otherwise, select the smaller region separated by ${\rm proj}(s)$ in each subarea as the correction $R_s$. Output the total correction set $R=\oplus_s R_s$.
    \item [6.]
    Initialize the measurement outcome of $X_{L}\otimes X_{L}$ (or $Z_{L}\otimes Z_{L}$) $m_o$ as the product of measurement results of the
    starred stabilizers in the $(d+3)/2$ layer. For path $s\in S$, if an odd number of syndrome changes of starred stabilizers in $s$ occurs before the $(d+3)/2+1$ layer, replace $m_o$ with $-m_o$. For $s$ that has an endpoint in $\{v_x^t|v_x\in V_R^m \cup V_R^2, t\leq (d+3)/2\}$, compare the syndrome of correction $R_s$ in the lattice $\mathcal{L}^{m}$ and the projections of actual syndrome changes. If they differ in an odd number of vertices, replace $m_o$ with $-m_o$. Output $m_o$ as the measurement outcome.
\end{itemize}

\subsection{Numerical results and further discussions}\label{further discussion}

\begin{figure*}[t]
\centering
\includegraphics[width=17.5cm]{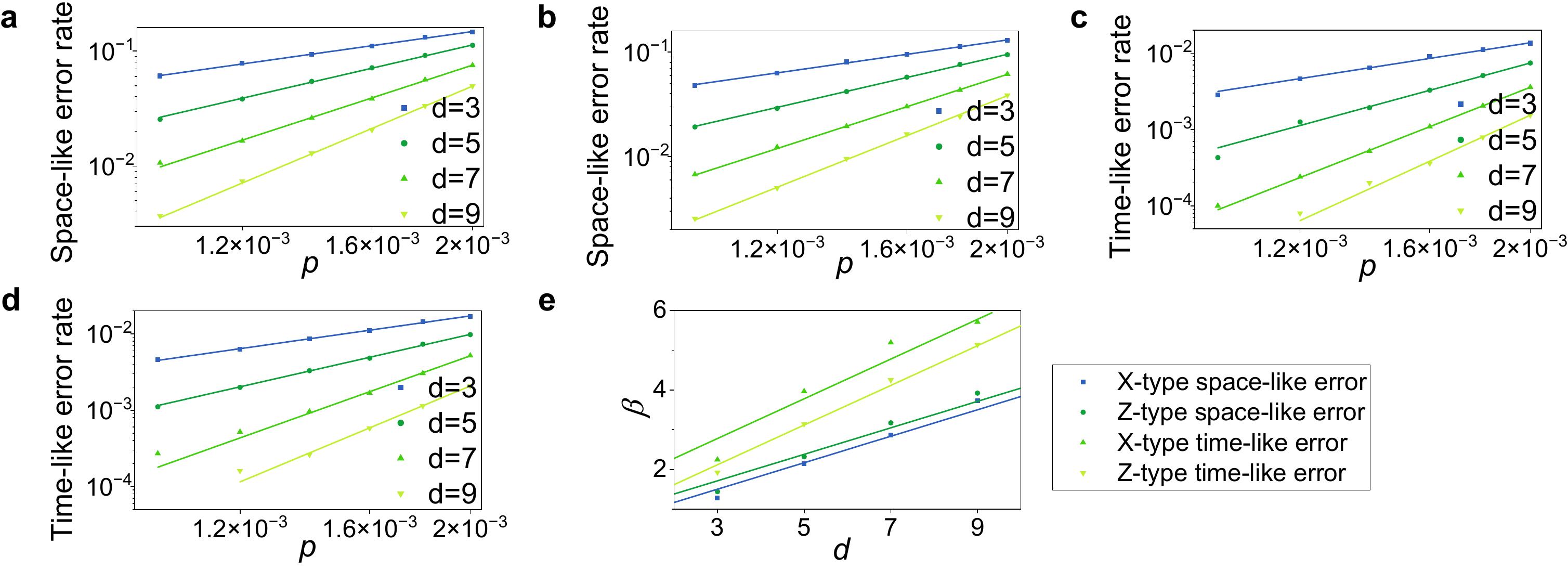}
\caption{\justify Space-like error rates and time-like error rates in the lattice surgery for various code distances $d$.  All the curves of error rates are fitted by the ansatz $\alpha p^\beta$. (a-b) $X$-type and $Z$-type space-like error rates. (c-d) $X$-type and $Z$-type time-like error rates. (e) Parameter $\beta$ of the curves in (a-d) for various code distances $d$. The parameters $\beta$ in space-like error rates and time-like error rates are fitted by $\beta=d/3+c_1$ and $\beta=d/2+c_2$ respectively.}\label{fig11}
\end{figure*}

We use Monte Carlo simulation to test the performance of our algorithm. For the results in the simulations, we say a space-like error occurs if $R$ causes a logical error, and a time-like error occurs if $m_o$ is an incorrect outcome and no space-like errors occur. The numerical results of the error rates of these two type of errors are shown in Fig.~\ref{fig11}. The error rate curves are also fitted by the ansatz $\alpha p^\beta$. In principle, we estimate that the curves space-like error rates have the parameters $\beta \sim d/3$, since our decoding algorithm is based on the projection decoder, which admits that it can correct all the errors of weight at most $d/3$ (up to an additive constant) \cite{beverland2021cost}. On the other hand, the time-like error curves are expected to have the parameters $\beta \sim d/2$, since they mainly result from the inaccurate measurement results of $d$ rounds of QEC circuits. Therefore, in Fig.~\ref{fig11}e, we use the lines with fixed slopes to fit the parameters $\beta$ of the curves, which indicates that the space-like effective code distance is less than that the time-like one. We also find that the time-like error rate is generally one to two orders of magnitude smaller than the space-like error rate, implying the dominated logical error in color code lattice surgery are space-like type. Note that the rates of $X$-type and $Z$-type errors are not equal in our simulation, and we speculate that this is due to the order of the $X$-type and $Z$-type stabilizer measurements.

It should be emphasized that, since our algorithm is based on the projection decoder \cite{delfosse2014decoding,beverland2021cost}, it may not be applicable to arbitrary forms of lattice surgery on color codes. Therefore, in the following, we will use the symmetry theory from Ref.~\cite{sahay2022decoder} to explain why our decoding algorithm can decode lattice surgery in the previous section and its applicable scope.

For a subset $\Sigma$ of the stabilizer group, we say it satisfies symmetry if $\prod_{S \in \Sigma}S_f = \mathbbm{1}$. It is easy to verify that any error will cause an even number of syndrome changes within $\Sigma$, thus allowing the matching algorithm to be applied to the restricted lattice corresponding to $\Sigma$. Now consider the set of stabilizers shown in Fig.~\ref{fig10}a. Define the tensor product of $X$ or $Z$ operators on all data qubits on the left (or right) boundary as the boundary operator $b_B$ ($b_G$). It is easy to verify the following relationships:
\[
\prod_{f \text{ is not green}}S_f \times b_B = \mathbbm{1},
\]
\[
\prod_{f \text{ is not blue}}S_f \times b_G = \mathbbm{1},
\]
and
\begin{equation}
\prod_{f \text{ is not red}}S_f \times b_G b_B = \mathbbm{1},
\end{equation}
where $f$ is the face in the primal lattice, and $S_f$ is $X$-type or $Z$-type stabilizer generator on the face $f$. Correspondingly, we can define three subsets of stabilizers with symmetry:
\[
\Sigma_{RB} = \{S_f \mid f \text{ is not green}\} \cup \{b_B\},
\]
\[
\Sigma_{RG} = \{S_f \mid f \text{ is not blue}\} \cup \{b_G\},
\]
and
\begin{equation}
\Sigma_{GB} = \{S_f \mid f \text{ is not red}\} \cup \{b_G, b_B\}.
\end{equation}
It is clear that $b_B$ and $b_G$ correspond to the boundary vertices $v_B$ and $v_G$ in Fig.~\ref{fig10}a, and these three subsets correspond to the three restricted lattices in the decoding algorithm. This explains why a matching algorithm-based decoder can effectively handle the lattice surgery example discussed in the previous section.

In general, more examples of lattice surgery are presented in Ref.~\cite{thomsen2022low} and Ref.~\cite{kesselring2024anyon}, which involve different types of boundaries between two regions. Ref.~\cite{kesselring2024anyon} classifies these boundaries into colored boundaries and Pauli boundaries. For lattices with only colored boundaries, finding the above symmetries is straightforward (they correspond to three restricted lattices). Therefore, decoding algorithms based on the MWPM (including our decoding algorithm) can effectively handle the decoding problems of this type of lattice surgery. However, for lattices with Pauli boundaries or mixed colored and Pauli boundaries, it is unclear whether such symmetries exist. The study of decoding algorithms on such lattices remains an open problem.

\section{State injection}\label{State injection}
In this section, we propose the state injection protocol of the triangular color code. First, we describe the process and the principles of the protocol in Sec~\ref{state injection protocol}. We also prove that the logical error rate of our protocol is lowest compared to the state injection protocol among all CSS codes. In Sec~\ref{post-selection schemes}, the performances of diverse post-selection schemes are shown by numerical results.

\subsection{State injection protocol}\label{state injection protocol}
\begin{figure}[t]
\centering
\includegraphics[width=6cm]{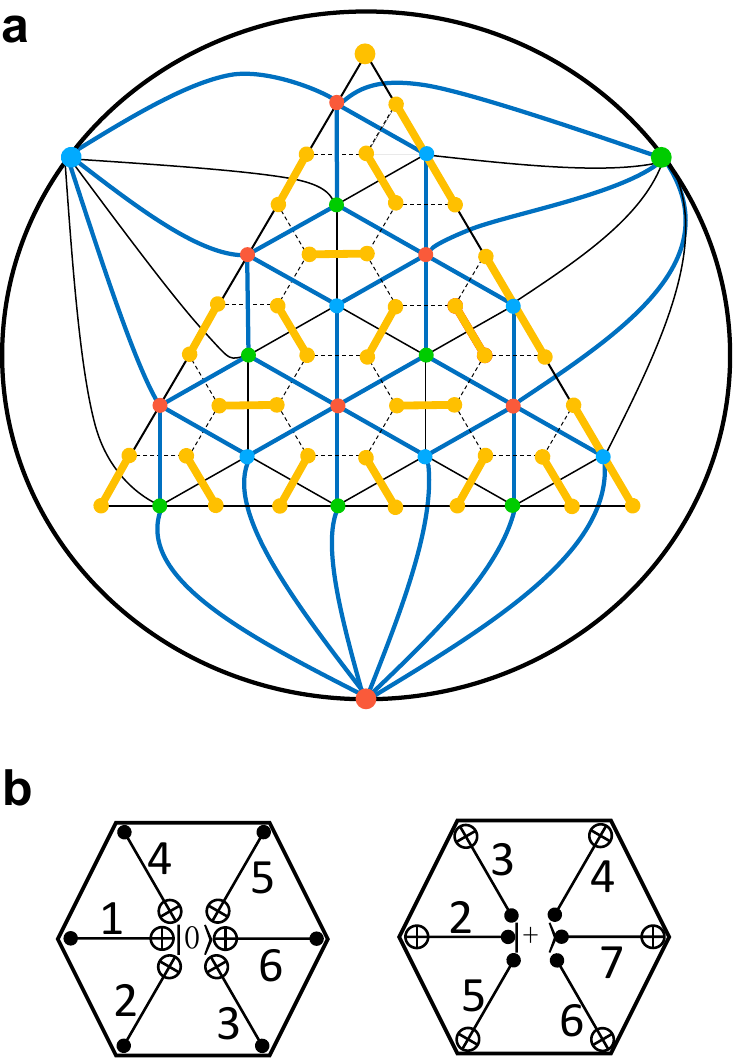}
\caption{\justify (a) Prime (dashed lines) and dual (solid lines) lattice of state injection protocol. The qubits are prepared in Bell pairs (yellow) except for the injected qubit on the top. The lattice is divided into several subareas by the blue lines and the outermost circle. (b) The CNOT schedule in the state injection protocol that offers the lowest logical error rate.}\label{fig13}
\end{figure}

State injection is the process of obtaining an arbitrary logical state $\ket{\psi_L}$ from a  physical state $\ket{\psi}$. Let us describe our state injection protocol of triangular color code. The protocol is executed in two steps. As illustrated in Fig.~\ref{fig13}a, in the first step, the top qubit is initialized to $\ket{\psi}$ and the other qubits are initialized to Bell states $\ket{\phi}=\frac{1}{\sqrt{2}}(\ket{00}+\ket{11})$ in pairs. In the second step, the QEC cycles are performed with the CNOT schedule we give in Fig.~\ref{fig13}b, which is carefully selected from 12 different schedules (see Appendix~\ref{ac} for details).

To prove that the final state is $\ket{\psi_L}$ after all noiseless operations, we can simply verify the following equation: 
\begin{equation}
\begin{aligned}
&\ket{\psi_L}\bra{\psi_L}\prod_i\frac{I+S_i}{2}\ket{\psi}\ket{\phi}^{\otimes m}\\
=&\frac{I+\alpha X_L + \beta Y_L + \gamma Z_L}{2}
\prod_i\frac{I+S_i}{2}
\ket{\psi}\ket{\phi}^{\otimes m}\\
=&(\prod_i\frac{I+S_i}{2})
\frac{I+\alpha X_L + \beta Y_L + \gamma Z_L}{2}
\ket{\psi}\ket{\phi}^{\otimes m}\\
=&\prod_i\frac{I+S_i}{2}
\ket{\psi}\ket{\phi}^{\otimes m},
\end{aligned}
\end{equation}
where we use $\ket{\psi}\bra{\psi}=(I+\alpha X + \beta Y + \gamma Z)/2$, $\ket{\psi_L}\bra{\psi_L}=(I_L+\alpha X_L + \beta Y_L + \gamma Z_L)/2$ and $S_i$ are the stabilizer generators of the triangular color code. Hence, the final state $\prod_i (I+S_i)/2\ket{\psi}\ket{\phi}^{\otimes m}$ is the $+1$ eigenstate of $\ket{\psi_L}\bra{\psi_L}$, i.e., $\prod_i(I+S_i)/2\ket{\psi}\ket{\phi}^{\otimes m}=\ket{\psi_L}$.

The state injection protocol is non-fault-tolerant because a single error in the circuit may cause a logical error that cannot be corrected by the decoding algorithm. More seriously, some errors cannot even be detected, which means post-selection is also powerless against
these errors. For example, the error occurring in the top qubit $\ket{\psi}$ before the first QEC cycle cannot be identified, since the injection protocol is applicable to an arbitrary state, even though it is a faulty state.

Through exhaustive search, we find that, in our protocol with post-selections, there are two types of errors that cannot be detected. The first is the single-qubit error when preparing the physical state $\ket{\psi}$. The second is the two-qubit errors after some CNOT gates in the first QEC cycle, including $X\otimes I$, $Z\otimes Z$ and $Y\otimes Z$ errors after the CNOT gate between
$\ket{\psi}$ and syndrome qubit $\ket{+}$, and $X\otimes X$ error after the CNOT
gate between $\ket{\psi}$ and syndrome qubit $\ket{0}$. We will discuss in detail later why these errors are undetectable.

Therefore, with the post-selection, the logical error rates of our state injection protocol is 
\begin{equation}
\begin{aligned}
P_L=\frac{4}{15}p_2+p_I+\frac{2}{3}p_1+\mathcal{O}(p^2),
\end{aligned}
\end{equation}
where $p_1$, $p_2$ and $p_I$ is the error rate of single-qubit gates, CNOT gates and initializing $\ket{0}$ or $\ket{+}$ state respectively, which is consistent with Ref.~\cite{li2015magic} and \cite{lao2022magic}. We assume that $\ket{\psi}$ initialization is performed one time step before the first CNOT gate on $\ket{\psi}$ is applied. Hence there is no need to consider the contribution of idling errors of $\ket{\psi}$ to $P_L$. 

As mentioned, the error rate of input state of magic state distillation will remarkably affect the output error rates. In a rough version of the injection protocol in Ref.~\cite{beverland2021cost}, the logical error rate after state injection is $6.07p+\mathcal{O}(p^2)$ and $4p$ is from single-qubit errors. If the error rate of a two qubit gate is assumed to be ten times as the rate of a single-qubit error and let $p_2=p$, the parameter $P_L$ in their protocol can be estimated as $P_L=(2.07+0.4)p+\mathcal{O}(p^2)$ and $P_L$ in our protocol is $4p/15+p/6+\mathcal{O}(p^2)$. Taking the 15 to 1 distillation protocol as an example, we use $p_o=35{P_L}^3$ to estimate the error rate of the distilled magic state. The output error rate $p_o$ from our state injection is over two orders of magnitude lower than that of the rough injection protocol. It is important to emphasize that in the comparison, we neglected the higher-order terms of $P_L$. This was done for two reasons: firstly, only the first-order term of $P_L$ is provided in Ref.~\cite{beverland2021cost}, and secondly, our numerical results indicate that the contribution of higher-order terms to $P_L$ is negligible for $p=10^{-4}$ and below.

Compared to previous works in surface codes, our protocol admits a lower logical error rate (see second column in Table.~\ref{t1}).  Further, we claim that the logical error rate of our protocol is lowest for state injection protocols among all CSS codes. Let us clarify it clearly with the following theorem.
\begin{theorem}
For any CSS code with standard stabilizer measurement circuits, the logical error rate of a state injection protocol is at least ${4}/{15}p_2+p_I+{2}/{3}p_1+\mathcal{O}(p^2)$ under the circuit-level depolarizing noise model.
\end{theorem}

Here standard stabilizer measurement circuits are the circuits using CNOT gates, measurement in $Z$- (or $X$-) basis and $\ket{0}$ (or $\ket{+}$)
as ancilla for $Z$-type (or $X$-type)  stabilizer measurements. Now let us prove the theorem by stating that several errors must not be detected in the state injection circuits. Without loss of generality, we suppose that $\ket{\psi}$ first couples with syndrome qubit $\ket{+}$ and then couples with syndrome qubit $\ket{0}$ in a QEC cycle. An obvious fact is that any error of $\ket{\psi}$ before the first stabilizer measurement is undetectable since $\ket{\psi}$ is arbitrary for an injection protocol. Therefore, the error of $\ket{\psi}$ initialization cannot be detected with rate $p_I+{2}/{3}p_1+\mathcal{O}(p^2)$. Besides, the $X$ error of $\ket{\psi}$ before the first CNOT gate between $\ket{\psi}$ and $\ket{0}$ but after the first CNOT between $\ket{+}$ and $\ket{\psi}$ is undetectable with the error rate of at least $p_2/15+\mathcal{O}(p^2)$, in which case the error is exactly $I\otimes X$ error after the first CNOT gate between $\ket{\psi}$ and $\ket{+}$. Further, utilizing the commuting relationship, a $Z\otimes Z$ error after the first CNOT gate between $\ket{+}$ and $\ket{\psi}$ is equivalent to an $I\otimes Z$ error before the stabilizer measurement, which is undetectable. Likewise, an $X\otimes X$ error after the first CNOT gale between $\ket{\psi}$ and $\ket{0}$ is also undetectable. Lastly, since $I\otimes X$ and $Z\otimes Z$ are undetectable, their product $Z\otimes Y$ after the first CNOT gate is also undetectable. In total, the logical error rate in a state injection
protocol is at least ${4}/{15}p_2+p_I+{2}/{3}p_1+\mathcal{O}(p^2)$, which is exactly the result in our protocol. Hence, we have demonstrated that the logical error rate in our protocol serves as a lower bound for the logical error rate achievable by any state injection protocol among CSS codes.

\subsection{Post-selection schemes and numerical results}\label{post-selection schemes}

\begin{figure*}[t]
\centering
\includegraphics[width=17.5cm]{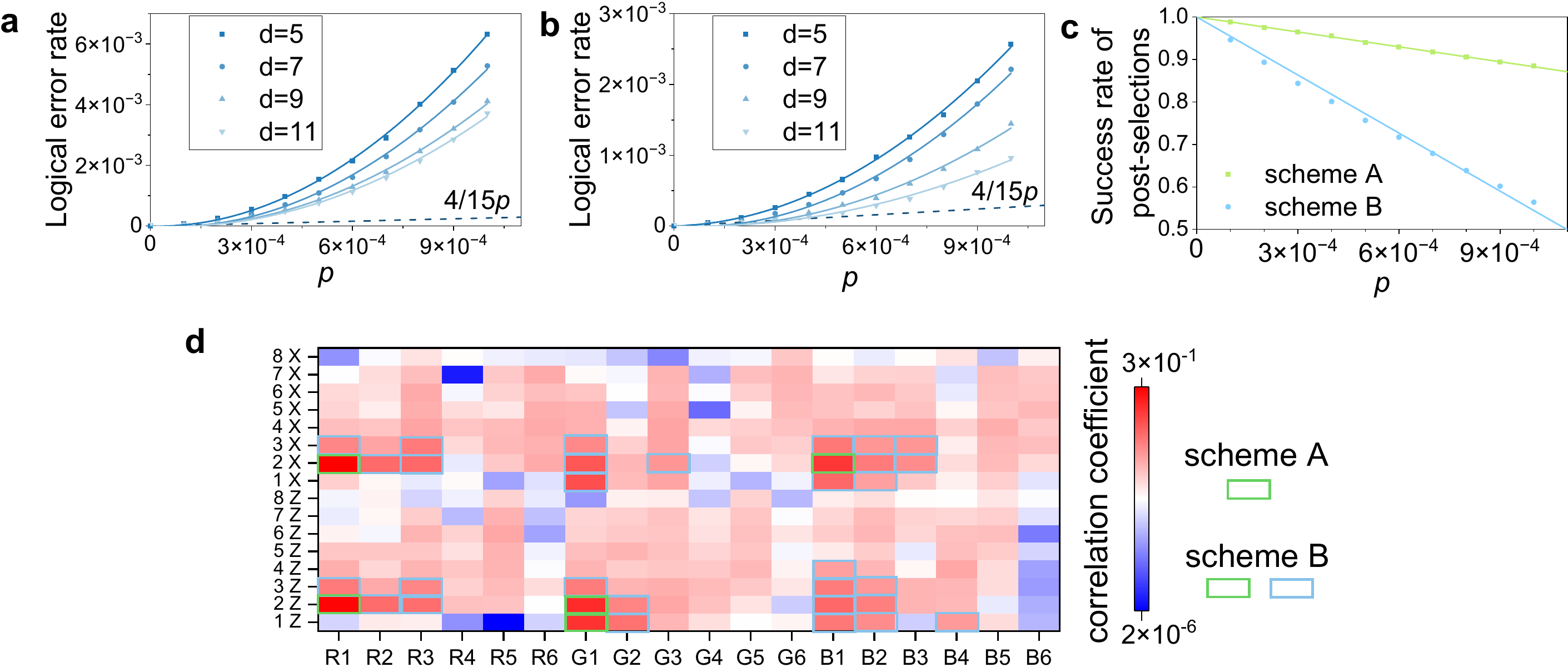}
\caption{\justify (a-b) The logical error rates of the state injection protocol in schemes A and B respectively, where the preparation of the injected state is assumed to be noiseless since it always causes a logical error. (c) Success rates of the post-selection as the linear function of the physical error rate $p$. The data are obtained from the state injection with $d=11$, and the results with other code distances are very close to them. (d) Correlation coefficients between syndrome changes and the occurrence of logical errors. The label, say R2 and 3Z, means the syndrome of the second red $Z$-type stabilizer (in the order from top to bottom and left to right in Fig.~\ref{fig13}a) in the third QEC round. The post-selected syndromes in schemes A and B are framed, respectively.}\label{fig14}
\end{figure*}

\begin{table*}[tbp]
\begin{center}
\renewcommand\arraystretch{1.5}
\begin{tabular}{p{40mm}ccp{40mm}}
\hline
\textbf{Injection protocol} & \quad \textbf{Logical error rate} \qquad \quad &  \textbf{Success rate of post-selection} \quad & \textbf{Other features}\\ \hline
on planar surface codes \cite{li2015magic} & $\frac{2}{5}p_2+2p_I+\frac{2}{3}p_1+\mathcal{O}(p^2)$ & $60\%$ with $p=10^{-3}$ & two steps where physical state is first injected to a $d=7$ logical qubit \\
on rotated surface codes from corner \cite{lao2022magic} & $\frac{3}{5}p_2+2p_I+\frac{2}{3}p_1+\mathcal{O}(p^2)$ & no data &  only gives results of $d\leq 5$\\
on rotated surface codes from middle \cite{lao2022magic} & $\frac{3}{5}p_2+p_I+\frac{2}{3}p_1+\mathcal{O}(p^2)$ & no data & only gives results of $d\leq 5$\\
on triangular color codes (this work)& $\frac{4}{15}p_2+p_I+\frac{2}{3}p_1+\mathcal{O}(p^2)$ & \thead{around $98\%$ with $p=10^{-4}$ \\and $88\%$ with $p=10^{-3}$}& one step to the logical qubit with any large code distance\\ 
\hline
\end{tabular}
\\[0pt]
\end{center}
\caption{Comparison of several state injection protocols}\label{t1}
\end{table*}

In the previous section, we theoretically analyze the leading order term of the
logical error rate in our protocol. In this section, we show the details of the post-selection schemes and estimate the logical error rates in diverse schemes by numerical simulations. In our simulations, the noise model in Sec~\ref{circuit-level noise model} continues to be used in the QEC circuits and the Bell state initialization circuits. We also assume that the initialization of $\ket{\psi}$ is perfect, since errors in this step always cause logical errors. The logic error rates are counted after $d+1$ rounds of QEC circuits.

Although the state injection protocol is non-fault-tolerant, it still requires decoding to suppress errors. We mainly follow the decoding strategy in Sec~\ref{decoding algorithm} but make three minor changes. First, in the 1st step when constructing the decoding graphs, we take the errors in the Bell state initialization circuits into account. Second, the definition of boundary vertex set $\Tilde{V}$ is modified as $\{v_x^{t}|v_x\in\{v_R,v_B,v_G\},t=1,2,...,d+1 \,\, {\rm or}\,\, v_x \,\,{\rm{is\,\, red}\,\,and}\,\,t=1\}$. This means that the red vertices in the first time layer are boundary vertices because the outcomes of the stabilizer measurements in these positions are undetermined. Third, in the 5th step, the path $s$ may start or end with red vertex in the first time layer, whose projection cannot divide $\mathcal{L}^*$ into two parts. In this case, we use the method similar to that in Sec~\ref{Decoding lattice} to divide $\mathcal{L}^*$ into several subareas to address this problem (see Fig.~\ref{fig13}a).

In order to further suppress the logical error rate $P_L$, post-selection schemes are introduced. The post-selection means that if some syndromes are not as expected, we discard current states and restart the state injection protocol. Our results show that the post-selections only need to be applied to five syndromes to minimize the leading order term of $P_L$. The post-selection scheme applied to these five syndromes is defined as Scheme A (see Fig.~\ref{fig14}). Through exhaustive search, we find that there are 90 errors (excluding single-qubit errors in preparing the $\ket{\psi}$ state)  will lead to a logical error with the probability of $12.4p+\mathcal{O}(p^2)$, and 86 of them can be detected by those five syndromes. The numerical results show that the logical error rate of post-selection scheme A is close to $4/15p$ when $p=10^{-4}$ (see Fig.~\ref{fig14}a).

When $p=10^{-3}$, however, the logical error rate is more than a dozen times of $4/15p$, since higher order terms contribute a lot in this case. Therefore, it is necessary to apply post-selection to more syndromes to suppress higher-order errors. To understand which syndromes are related to logical errors, we test the correlation coefficients of syndrome changes and the occurrence of logical errors by Monte Carlo simulations in a $d=7$ triangular color code with $p=10^{-3}$ (see Fig.~\ref{fig14}d). The correlation coefficient of events $A$ and $B$ is computed by
\begin{equation}
\begin{aligned}
\rho_{AB}=\frac{|p(AB)-p(A)p(B)|}{\sqrt{p(A)(1-p(A))p(B)(1-p(B))}},
\end{aligned}
\end{equation}
where $p(A)$, $p(B)$, $p(AB)$ are the probabilities of $A$, $B$, both $A$ and $B$ occurring respectively. 

According to these correlation coefficients, one can design the post-selection scheme flexibly with the syndromes whose correlation coefficients are larger. Apparently, the post-selection with more syndromes will lead to lower success rate of post-selection. Here we present an example with 35 post-selected syndromes (scheme B) where the logical error rate is reduced by 4 times when $p=10^{-3}$ compared to scheme A, while the success rate of post-selection is still over $56\%$. More examples of post-selection schemes with lower success rates are shown in Appendix~\ref{ac}.

Contrary to previous works on state injections in surface codes, the success rate of post selections in our protocol does not decay sharply as the code distance increases, but is only linearly related to the physical error rate (see Fig.~\ref{fig14}c and Table.~\ref{t1}). On account of the fact that errors farther from the top qubit $\ket{\psi}$ are on a longer logical string (i.e., the data qubit string that connects left and right boundaries), the stabilizers of the post-selected syndromes are located around the top qubit. Therefore, the number of post-selected syndromes does not increase with the code distance, which means one can inject a physical state into a logical state with arbitrary code distance in one step, rather than two steps as in Ref.~\cite{li2015magic}. In addition, this feature provides our protocol with a high success rate of post-selection, around $98\%$ 
($88\%$) in scheme A when $p=10^{-4}$ ($10^{-3}$).

\section{Conclusion and outlook}\label{Conclusion and outlook}
In this paper, we solved several key problems in the road towards practical fault-tolerant quantum computation with 2-D color codes, including improving the threshold of the triangular color code, decoding color code lattice surgery with colored boundaries under circuit-level noise and proposing an optimal state injection protocol. We believe these works are expected to promote the development of quantum computing based on color codes.

Our results shows that more accurate matching weights can effectively improve the threshold of color code. A better weight setting is to calculate the posterior probability of errors in the circuits by belief propagation algorithm, known as BP-matching algorithm, whose improvement has been demonstrated in surface codes \cite{higgott2023improved,criger2018multi}. The color code threshold is expected to be further improved by the BP-matching algorithm. Besides, the logical error rates of the color code lattice surgery in this work are higher than the results of the mature surface coding protocols \cite{chamberland2022circuit,chamberland2022universal}. Nevertheless, we believe that the results can be further improved by optimizing the circuits in lattice surgery protocol, as Ref.~\cite{gidney2023new} narrowing the performance gap between color codes and surface codes in storage experiments through circuit optimization.

There are still many other challenges of quantum computing based on color code left for future work. For example, the PBC model is probably not the final form of quantum computation based on color codes, since it does not utilize the advantages of transversal Clifford gates on color codes. According to the specific quantum algorithm, a proper way to use color codes for quantum computing still needs to be developed. 

In addition, we also note the color codes with other boundaries, such as square color codes or thin color codes \cite{kesselring2024anyon,thomsen2022low}, which are also the candidates for the color code quantum computing. These color codes encode more than one logical qubit in an individual patch and may reduce the overhead further under the special structured noise (e.g., biased noise). When the boundaries are of the colored type, we do not perceive fundamental difficulties in applying our lattice surgery decoding strategies to them. However, if Pauli boundaries are considered, decoding lattice surgery on such boundaries remains an unresolved issue. In particular, thin color codes with $d_z>d_x$ can reduce the cost of magic state distillation, since the distillation error is more sensitive to the $Z$-type error \cite{litinski2019magic}. Based on the feature of parallel measurements and our lattice surgery decoding algorithm, estimating and optimizing the time and space overhead of magic state distillation in a color-code-based quantum computation will also be an interesting question.

Lastly, our state injection protocol can enhance recent results on magic state preparation \cite{chamberland2020very}, further improving the fidelity of magic states. A possible future direction is the quantum computation with surface and color codes combined, where surface codes are used for computation, and color codes are used for distillation.

\acknowledgments
This work was supported by the National Natural Science Foundation of China (Grant No. 12034018) and Innovation Program for Quantum Science and Technology (Grant No. 2021ZD0302300).

\appendix
\section{Automated procedure for constructing decoding graphs}\label{ab}
The automated procedure extracts error information of noisy circuits based on the circuit-level depolarizing noise model. In a noisy quantum circuit, we approximate every noisy operation by an ideal operation followed by the depolarizing error channel $\mathcal{E}_{1}$ or $\mathcal{E}_{2}$. We simulate the circuits in the Heisenberg representation~\cite{gottesman1998heisenberg}, randomly inserting Pauli errors with corresponding probability. Since the circuits are Clifford circuits, according to the Gottesman-Knill theorem, they can be efficiently simulated in polynomial time~\cite{nielsen2010quantum,PhysRevA.70.052328}. 

As mentioned in the main text, in each time, the automated procedure simulates an ideal QEC circuit attached to a single error in the different channels in the circuit. If the error is located in channel $\mathcal{E}_{1}$, the error is $X$, $Y$ or $Z$. If the error is located in channel $\mathcal{E}_{2}$, the error is one of $P_1\otimes P_2$ where $P_1,P_2\in{I,X,Y,Z}$ and $P_1\otimes P_2\neq I \otimes I$. 

In fact, there is no need to simulate every round of the QEC cycle, since the effect of a single error is limited to two the difference between two layers. For each position and type of error, we only simulate the circuit in two QEC rounds, where the error is only added in the first round and the circuit in the second round is perfect. We record the vertex sets corresponding to the syndrome changes in each simulation.

In the automated procedure, we connect the syndrome changes in pairs to form edges by the following rule. First, we connect the vertices that are projected into the same position, if they exist. Then we connect the remaining vertices with the same color, if they exist. Using the CNOT schedules in Fig.~\ref{fig6}, there are at most two vertices remaining unmatched in each decoding graph after this step. If there are two vertices left, we connect them and if there is only one vertex left, we check the syndrome of the error and connect the vertex with the proper boundary vertex.

Based on these connections, the edges in the first two layers of $\mathcal{G}_{C}$ are determined, denoted by $e_0=(v_1^{t_1}, v_2^{t_2})$. All edges $e_0$ obtained from the simulations in two QEC rounds form an edge set $\beta_0$. Then one can construct the edge set $\beta$ in $\mathcal{G}_{C}$ by $\beta_0$:
\begin{equation}
\begin{aligned}
\beta=\{e=(v_1^{t_1}, v_2^{t_2})|v_1^{t_1}, v_2^{t_2}\in\mathcal{G}_{C}, e//e_0,e_0\in \beta_0\},
\end{aligned}
\end{equation}
where we say two edges $e_1=(v_1^{t_1}, v_2^{t_2})$, $e_2=(v_3^{t_3}, v_4^{t_4})$ are parallel in a 3-D lattice if they have the same vertical projection and $t_3-t_1=t_4-t_2$, notated by $e_1//e_2$.

Suppose edge $e_0\in\beta_0$ is obtained from the syndrome changes of the single error $\epsilon$, the error rate corresponding to $e_0$ is $p_{\epsilon}(e_0)$, where $p_{\epsilon}(e_0)=p/3$ if ${\epsilon}$ is the result of $\mathcal{E}_{1}$, otherwise $p_{\epsilon}(e_0)=p/15$. By accumulating $p_{\epsilon}(e_0)$ with different $\epsilon$, we obtained the error-rate-related weight of the edge $e=(v_1^{t_1}, v_2^{t_2}) \in \beta$:
\begin{align}
& w_e=
\left\{
\begin{aligned}
&-\log\sum_{\substack{ e_0\in\beta_0^{(1)},\\ {\rm and}\,\,e_0//e }}
\sum_{{\epsilon}} p_{\epsilon}(e_0), \quad{\rm if}\quad t_1=t_2=1,\\
&-\log\sum_{\substack{ e_0\in\beta_0^{(2)},\\ {\rm and}\,\,e_0//e }}
\sum_{{\epsilon}} p_{\epsilon}(e_0), \quad{\rm if}\quad t_1=t_2=d+1,\\
&-\log\sum_{\substack{e_0\in\beta_0,\\ {\rm and}\,\,e_0//e }}
\sum_{{\epsilon}} p_{\epsilon}(e_0), \quad\quad{\rm otherwise},
\end{aligned}
\right.
\end{align}
where $\beta_0^{(1)}=\{e_0|e_0=(v_i^{1}, v_j^{1}) \in \beta_0\}$ and $\beta_0^{(2)}=\{e_0|e_0=(v_i^{2}, v_j^{2}) \in \beta_0\}$. Note that the edges in the time boundaries (i.e., the first and last layers) are considered individually in the first two cases.

\section{Different CNOT schedules and more post-selection schemes in the state injection protocol}\label{ac}
\begin{figure*}[htb]
\centering
\includegraphics[width=17.5cm]{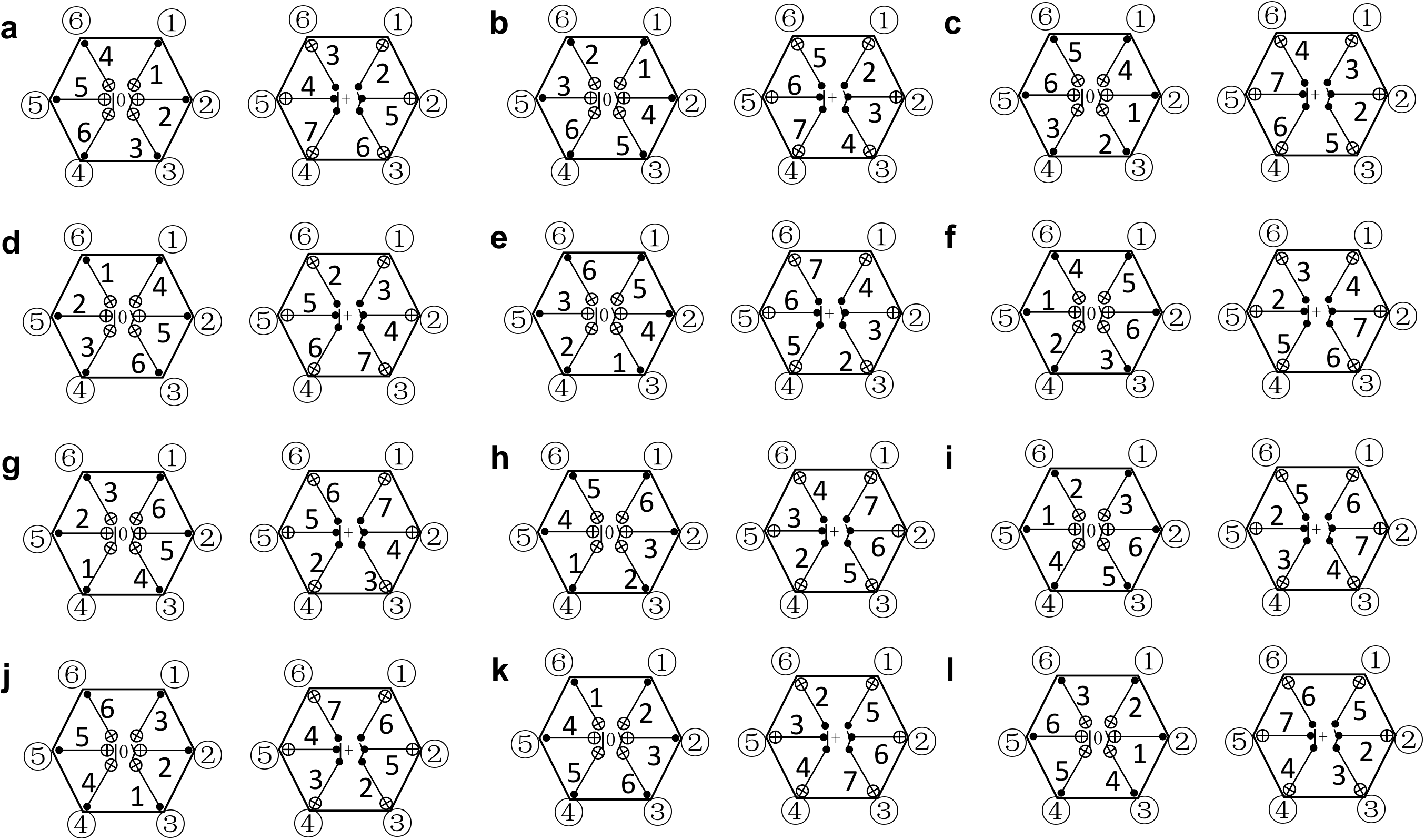}
\caption{\justify Details of the 12 types of CNOT schedules we considered in the state injection. (a-l) correspond to $0^+$,$0^-$,$1^+$, ..., $5^-$, respectively. The CNOT schedules with $+$ are obtained by rotating $0^+$, and CNOT schedule $k^-$ is the  flip of $k^+$.}\label{fig15}
\end{figure*}

In order to find the optimal CNOT schedule of state injection, we test 12 kinds of CNOT schedules as shown in Fig.~\ref{fig15}. Note that $0^+$ is the CNOT schedule used in color code QEC circuits and other schedules are obtained by flipping or rotating $0^+$. We use these CNOT schedules since the flip or rotation operation will hold the relationship in the criterion of a proper CNOT schedule. The complete data of logical error rates to the leading order are shown in Table.~\ref{t2}.

In Table.~\ref{t3}, we tested more post-selection schemes for the case of $p=10^{-3}$ and $d=11$. Recall that we prioritize syndromes with higher correlation coefficients for posterior selection. As more syndromes are post-selected, $P_L$ decreases to below $p$, with the cost of reduction in the success rate of post-selection. Note that the logical error rate tends to stabilize with more syndromes being post-selected. We infer that this is due to the inherent logical error rate of $d=11$ logical qubits, approximately $5\times10^{-4}$. Although post-selection can reduce this value, the effect is not significant because the range of stabilizers we post-select is limited. Therefore, as the code distance of logical qubits continues to increase, achieving lower logical error rates than those listed in Table.~\ref{t3} is possible.

\begin{table*}
\begin{center}   
\caption{Logical error rates in different CNOT schedules with and without post-selection}\label{t2} 
\renewcommand\arraystretch{1.5}
\begin{tabular}{|c|c|c|}   
\hline   \textbf{CNOT schedule} & \thead{\textbf{$P_L$ with post-selection}\\ \textbf{(to the leading order)}} & \thead{\textbf{$P_L$ without post-selection} \\ \textbf{(to the leading order)}}\\
\hline   \textbf{$0^+$} & $0.6p$ & $10.6p$\\
\hline   \textbf{$0^-$} & $1.2\dot{6}p$ & $13.2\dot{6}p$\\
\hline   \textbf{$1^+$} & $0.\dot{3}p$ & $9.8\dot{6}p$\\
\hline   \textbf{$1^-$} & $0.5\dot{3}p$ & $11.1\dot{3}p$\\
\hline   \textbf{$2^+$} & $0.\dot{3}p$ & $10.8p$\\
\hline   \textbf{$2^-$} & $0.2\dot{6}p$ & $12.4p$\\
\hline   \textbf{$3^+$} & $0.6p$ & $11.5\dot{3}p$\\
\hline   \textbf{$3^-$} & $1.2\dot{6}p$ & $12.2p$\\
\hline   \textbf{$4^+$} & $1.2p$ & $12.0\dot{6}p$\\
\hline   \textbf{$4^-$} & $p$ & $12.\dot{6}p$\\
\hline   \textbf{$5^+$} & $0.9\dot{3}p$ & $13.6p$\\
\hline   \textbf{$5^-$} & $p$ & $11.7\dot{3}p$\\
\hline
\end{tabular}  
\end{center}   
\end{table*}

\begin{table*}
\begin{center}   
\caption{Logical error rates with other post-selection schemes}\label{t3} 
\renewcommand\arraystretch{1.5}
\begin{tabular}{|c|c|c|}   
\hline   \thead{\textbf{Number of post-selected }\\ \textbf{syndromes}} & \thead{\textbf{Logical error rate $P_L$}} & \thead{\textbf{Success rate of} \\ \textbf{the post-selection}}\\
\hline   \textbf{$50$} & $8.9\times10^{-4}$ & $40\%$\\
\hline   \textbf{$62$} & $8.2\times10^{-4}$ & $34\%$\\
\hline   \textbf{$76$} & $7.7\times10^{-4}$ & $27\%$\\
\hline   \textbf{$93$} & $6.0\times10^{-4}$ & $21\%$\\
\hline   \textbf{$124$} & $5.7\times10^{-4}$ & $14\%$\\
\hline   \textbf{$136$} & $5.4\times10^{-4}$ & $12\%$\\
\hline   \textbf{$148$} & $5.2\times10^{-4}$ & $10\%$\\
\hline   \textbf{$164$} & $5.2\times10^{-4}$ & $9\%$\\
\hline
\end{tabular}  
\end{center}   
\end{table*}

\section{Other details of the simulations}\label{ad}
We use Monte Carlo simulation to estimate the logical error rates in different cases. In each time of simulation, QEC circuits are executed for $d+1$ rounds in the color code decoding and state injection, and $2d+1$ rounds (including the merged step and split step) in the lattice surgery. In both three situations, the circuit in the last round is assumed to be noiseless. 

Another key question is how to determine if there is a logical error after decoding. In the color code decoding, we check whether the product of correction $R$ and actual error $\epsilon$ is a logical operator (up to a stabilizer) by commutation relations. 

In the lattice surgery, space-like and time-like errors should be considered separately. Note that the space-like logical error must anticommute with an odd number of starred stabilizers. The output correction $R$ may cause a logical error as well as the stabilizers of several Bell states in the middle region, since some boundary vertices are located inside the lattice. These stabilizers are trivial for our decoding, because they always anti-commute with an even number of starred stabilizers.  The space-like errors in the top and bottom logical qubits are equivalent since $X_L\otimes X_L$ and $Z_L\otimes Z_L$ have been measured. For the time-like error, we set all the red boundary vertices with the initial syndromes $+1$ to guarantee that the expected outcome of $X_L\otimes X_L$ or $Z_L\otimes Z_L$ measurement is $+1$. Then we check the actual outcome to determine if there is a time-like logical error. 

In the state injection, the correction $R$ may differ from the actual error $\epsilon$ by $X\otimes X$ (or $Z\otimes Z$) error in the qubits of the initial Bell pairs whose syndrome corresponds to two red vertices, since the measurements of red stabilizers do not have determined initial syndromes. However, they will be determined fault-tolerantly after sufficient rounds of measurements and then $X\otimes X$ will be corrected. Therefore, we do not think the decoding fails when such errors are left after decoding. From another perspective, such $X\otimes X$ (or $Z\otimes Z$) errors will not change the measurement result of the logical Pauli operators ($X_L$, $Y_L$ or $Z_L$) since the Bell pairs and the logical Pauli operator always have an even number of common qubits.

Lastly, we list the simulation times of the numerical results in the main text. In Fig.~\ref{fig8}a
and Fig.~\ref{fig11}, each point is obtained by over $10^6$ simulations. In Fig.~\ref{fig14}a and Fig.~\ref{fig14}b, we obtain each point by $10^3/p$ times of simulations, where $p$ is the physical error rate ranging from $10^{-4}$ to $10^{-3}$.

\section{Direct comparison with the threshold in Ref.~[33]}\label{dc}

\begin{figure*}[tb]
\centering
\includegraphics[width=16cm]{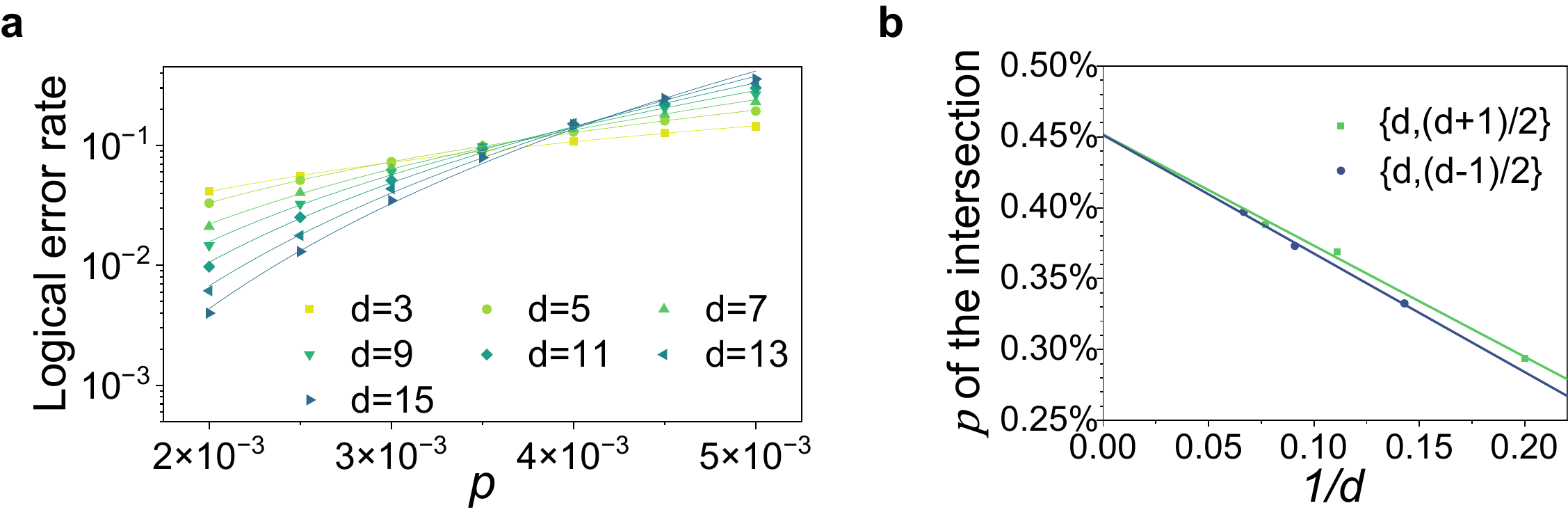}
\caption{\justify (a) Logical error rates $P_L$ of the triangular color code under the same circuit-level noise model as that in Ref.~\cite{beverland2021cost} with \(d \leq 15\).  (b) The intersections of pairs of curves of $P_L$ with distances $\{d,(d+1)/2\}$ and $\{d,(d-1)/2\}$. From the linear extrapolation of the data, the threshold is around $0.45\%$.}\label{fig16}
\end{figure*}

In the main text, we avoided directly comparing our threshold results with those in Ref.~\cite{beverland2021cost} because our circuit-level noise model is slightly weaker than theirs. Specifically, they assume measurement results flip with probability \(p\), whereas we assume \(2p/3\) (the equivalent result for a depolarizing channel $\mathcal{E}_{1}$). Here, we provide some simulation results under the same noise model as in Ref.~\cite{beverland2021cost}. We tested the logical error rates of the triangular color code for \(d \leq 15\) and evaluated the threshold using the same method as in the main text (see Fig.~\ref{fig16}). The results show that both the logical error rates and the threshold (around 0.45\%) outperform those in Ref.~\cite{beverland2021cost} (around 0.37\%), directly demonstrating the improvements our decoding algorithm provides.

\newpage
\bibliographystyle{unsrt}
\bibliography{ref}
\end{document}